\documentclass[a4paper,number,5p,twocolumn]{elsarticle}

\usepackage[english]{babel}
\usepackage[utf8]{inputenc}
\usepackage{graphicx}
\usepackage{setspace}
\usepackage[usenames,dvipsnames]{xcolor}
\usepackage{lineno}
\usepackage{ulem}
\usepackage{amsfonts}
\usepackage{amsmath}
\usepackage{amssymb}
\usepackage{bm}
\usepackage{soul}
\usepackage{multirow}
\usepackage[mathscr]{euscript}
\usepackage{enumerate}
\usepackage[]{hyperref}
\usepackage{wasysym}

\mathchardef\mhyphen="2D
\DeclareMathOperator*{\assembly}{\mathrm{A}}

\newcommand{\Ckubc}{\textrm{KU}}		
\newcommand{\Cminkbc}{\textrm{mK}}		
\newcommand{\Fref}[1]{Fig.~\ref{#1}}
\newcommand{\Tref}[1]{Tab.~\ref{#1}}	
\newcommand{\Eref}[1]{Eq.~(\ref{#1})}

\newcommand{\Sref}[1]{Section~\ref{#1}}
\newcommand{\Aref}[1]{\ref{#1}}
\newcommand{\norm}[1]{\|#1\|}
\newcommand{\loc}{^\ell}

\newcommand{\nelem}{n_{e}}						
\newcommand{\nboundarynodes}{n^{\boundary}}		
\newcommand{\nallnodes}{n^{\textrm{tot}}}		
\newcommand{\nboundarybeams}{n_{e}^{\boundary}}	
\newcommand{\etal}{~et~al.}
\newcommand{\Matlab}{Matlab}
\newcommand{\Alporas}{Alporas\textsuperscript{\textregistered}}
\newcommand{\alporas}{Alporas\textsuperscript{\textregistered}}
\newcommand{\RVE}{RVE}
\newcommand{\bmath}[1]{\ensuremath{\bm{#1}}}
\newcommand{\de}[1]{\,{\mathrm d}#1}

\newcommand{\atx}{\ensuremath{(\tens{x})}}
\newcommand{\half}{\frac{1}{2}}
\newcommand{\set}[1]{{\mathbb #1}}

\newcommand{\setR}{\set{R}}

\newcommand{\domain}{\Omega}
\newcommand{\boundary}{\partial\domain}

\newcommand{\measure}[1]{|#1|}

\newcommand{\scal}[1]{\mathnormal{#1}}
\newcommand{\tens}[1]{\boldsymbol{#1}}				
\newcommand{\tenss}[1]{\bmath{#1}} 					
\newcommand{\tensf}[1]{\bmath{\mathbf{#1}}} 		
\newcommand{\ex}{\tens{x}} 							
\newcommand{\trn}{{\sf T}}

\newcommand{\semtrx}[1]{\mathsf{#1}} 					
\newcommand{\sevek}[1]{\mathsf{#1}} 					
\newcommand{\scontr}{\cdot}
\newcommand{\dcontr}{\,\colon}

\newcommand{\Strain}{E}

\newcommand{\strain}{\varepsilon}

\newcommand{\stress}{\sigma}
\newcommand{\Stress}{\Sigma}

\newcommand{\potential}{J}
\newcommand{\D}{D}
\newcommand{\homg}{\mathrm{hom}}
\newcommand{\Dhom}{\tensf{\D}^{\homg}}
\newcommand{\Dhommtrx}{\semtrx{\D}^{\homg}}

\newcommand{\macrodensity}{\widehat{W}}
\newcommand{\microdensity}{W}

\newcommand{\avgs}[1]{\left\langle #1 \right\rangle} 	

\newcommand{\rem}[1]{}

\DeclareGraphicsExtensions{.pdf,.eps,.png,.jpg}	
\makeatletter
\def\ps@pprintTitle{
	\let\@oddhead\@empty
	\let\@evenhead\@empty
	\def\@oddfoot{{\small© 2016. This manuscript version is made available under the \href{http://creativecommons.org/licenses/by-nc-nd/4.0/}{ CC-BY-NC-ND 4.0 license.}}\hfill{}}%
	\let\@evenfoot\@oddfoot
}
\makeatother

\soulregister\cite7
\soulregister\ref7
\soulregister\pageref7

\journal{Computers \& Structures}

\begin{document}

\begin{frontmatter}

\title{A jigsaw puzzle framework for homogenization of high porosity foams\tnoteref{t1}}
\tnotetext[t1]{The author's post-print manuscript of the article published in \mbox{\textit{Computers \& Structures}}, \href{http://dx.doi.org/10.1016/j.compstruc.2016.01.003}{DOI: 10.1016/j.compstruc.2016.01.003}.}

\author[ctu]{Martin Do\v{s}k\'{a}\v{r}}
\ead{martin.doskar@fsv.cvut.cz}
\author[ctu]{Jan Nov\'{a}k\corref{cor}}
\ead{novakj@cml.fsv.cvut.cz}
\cortext[cor]{Corresponding author. Tel.:~+420-224-354-606}
\address[ctu]{Department of Mechanics, Faculty of Civil Engineering, Czech Technical University in Prague, Th\'{a}kurova 7, \mbox{166 29 Praha 6}, Czech~Republic}


\begin{abstract}
An approach to homogenization of high porosity metallic foams is explored. The emphasis is on the \Alporas{} foam and its representation by means of two-dimensional wire-frame models. The guaranteed upper and lower bounds on the effective stiffness coefficients are derived by the first-order homogenization with the uniform and minimal kinematic boundary conditions at heart. This is combined with the method of Wang tilings to generate sufficiently large material samples along with their finite element discretization. The obtained results are compared to experimental and numerical data available in literature and the suitability of the two-dimensional setting itself is discussed.
\end{abstract}



\end{frontmatter}

\section{Introduction}
\label{sec:introduction}

Highly porous metallic foams possess an extensive application potential. These materials feature high energy absorption, strength, and stiffness at very low weight, which makes them appealing for the automotive, aircraft, and defense industry, to name a few~\cite{ashby2000metal,banhart2001manufacture,hanssen2002closerange}.
In order to foster new application areas, a qualified understanding of the foam behavior and relevant predictive tools are required. Computer models, in particular, are regarded as a key ingredient to optimization of either/both the microstructure geometry or/and final products made of these materials~\cite{fullwood2010microstructure}.
A consensus regarding the approximation of the behavior of porous solids seems to exist. Open-cell foams\footnote{As Banhart~\cite{banhart2001manufacture} pointed out, sponge would be an appropriate term for open-cell highly porous solids.} are usually represented with the three-dimensional beam models, while their closed-cell counterparts require an addition of membrane elements acting as the cell walls~\cite{gibson1982mechanics}. Nevertheless, in the case of very thin walls even the behavior of closed-cell foams can be approximated with beams~\cite{banhart2001manufacture,despois2006uniaxial,ashby2000metal,koudelka2012effective}. Following this assumption, Ashby and Gibson presented a three-dimensional beam model of the unit cell and derived solutions to the overall thermo-mechanical parameters based purely on material porosity~\cite{gibson1982mechanics}.
Our goal is to explore the adequacy of a two-dimensional wired model discretized with beam elements. We build on the recent outcomes by N\v{e}me\v{c}ek\etal{}~\cite{nemecek2013twoscale}, who developed a planar beam model based on the approximation of foam ligament geometry by the Voronoi tessellation.
The objectives of the paper are threefold:
\begin{enumerate}[i)]
\item to investigate the influence of real geometry of foam ligaments and the proper size of computational models~\cite{nemecek2013twoscale};
\item to provide the upper and lower bounds on the effective stiffness coefficients for the two-dimensional beam model via the first-order homogenization procedure; 
\item to validate obtained results against the experimental and numerical results from~\cite{nemecek2013twoscale} and to question the assumption of the vanishing Poisson ratio made therein.
\end{enumerate}

The first objective is addressed with the help of Wang tilings, a concept recently introduced to Materials Engineering community~\cite{novak2012compressing}. In particular, the foam microstructure is compressed within a set of smaller domains, called Wang tiles. The morphology of the tiles is designed such that the tiles are microstructurally compatible across the corresponding edges. As a result the reconstructed material microstructure remains continuous across the gridline of the regular lattice to which the tiles are accommodated during synthesis of a computational model, see~\Fref{fig:tileillustration}(c). The method is extremely efficient in producing arbitrarily diverse ensembles of arbitrarily sized and geometrically consistent microstructure realizations in a fully stochastic setting~\cite{doskar2015aperiodic}.
Moreover, creating the finite element mesh on the level of tiles avoids expensive mesh generation of each microstructure realization. Altogether, we are able to reach for a proper size of the computational model, which is expected to be relatively large because of the infinite contrast in constituent properties~\cite{dirrenberger2014towards}.

As regards the second objective, we reformulate the first-order homogenization procedure for the wire-frame finite element models by means of macroscopic degrees of freedom~\cite{michel1999effective}. The upper bound on the apparent stiffness is obtained from the ensemble of microstructures exposed to the kinematic uniform boundary conditions (KUBC). The lower bound is rendered by applying the minimal kinematic boundary conditions (mKBC)~\cite{mesarovic2005minimal}. In order to compare our results with those reported in~\cite{ashby2000metal,banhart2001manufacture,miyoshi2000alporas,nemecek2013twoscale}, the bulk and shear moduli are derived from the homogenized stiffness matrices by assuming material isotropy.

The paper is structured as follows. In the next chapter \alporas{} aluminum foam, the material of interest, is characterized in brief. Section \ref{sec:strategy} covers the adopted modeling strategy addressing the first objective. In particular, Wang tilings are presented first together with their use in the context of numerical homogenization. Description of microstructure discretization into planar beams is also provided in this section. The first order homogenization is reviewed in~\Sref{sec:homogenization} with emphasis on the beam model and the boundary conditions connected to the upper and lower bounds on the effective stiffness. Numerical results are provided in~\Sref{sec:results} and discussed in the last section.

We use the following nomenclature within the text. Scalar quantities are denoted by plain letters, e.g. $\scal{a}$ or $\scal{A}$, matrices by bold sans-serif font, e.g. $\sevek{a}$ or $\semtrx{A}$, and tensorial quantities by bold serif letters, e.g. $\tenss{a}$ or $\tensf{A}$. In addition, we adopt the standard Voigt matrix representation of symmetric second- and fourth-order tensors.

\section{Alporas}
\label{sec:alporas}

\Alporas{} is the closed-cell aluminum foam commercially produced by Shinko Wire Company,~Ltd.~\cite{miyoshi2000alporas} with porosity between 90~\% and 92~\%~\cite{ashby2000metal}. The pore space is composed of approximately polyhedral voids with the mean size of around 3~mm. The main field of its application lies in the energy absorption structures. Predominantly, \alporas{} is used in production of self-supporting sound absorbers~\cite{miyoshi2000alporas}. The sandwich plates comprising of \alporas{} foam core were also reported to outperform common steel plates under blast loading in e.g.~\cite{radford2006response}.

Regarding the linear elastic behavior, the design guide to metallic foams~\cite{ashby2000metal} suggests the overall parameters enlisted in \Tref{tab:alporasproperties}. 
\begin{table}[ht!]
	\centering
	\caption{Properties of \Alporas{} foam \cite{ashby2000metal}. Here $K$, $G$ and $E$ stand for the bulk, shear and Young moduli, respectively, and $\nu$ is the Poisson ratio.}
	\label{tab:alporasproperties}
	\renewcommand{\arraystretch}{1.35}
	\begin{tabular}{cccc} \hline
		$K$ [GPa] & $G$ [GPa] & $E$ [GPa] & $\nu$ [-] \\ \hline
		0.9 - 1.2 & 0.30 - 0.35 & 0.4 - 1.0 & 0.31 - 0.34 \\\hline
	\end{tabular}
\end{table}
However, the region of linear elastic behavior is rather limited. Nonlinear or inelastic behavior develops soon after the loading onset due to wall buckling, contacts between ligaments, and local yielding, which makes interpretation of overall elastic parameters from experimental data a challenging task.

\begin{figure}[ht!]
  \centering
  \setlength{\tabcolsep}{1pt}
  \begin{tabular}{ccc}
    \includegraphics[height=2.85cm]{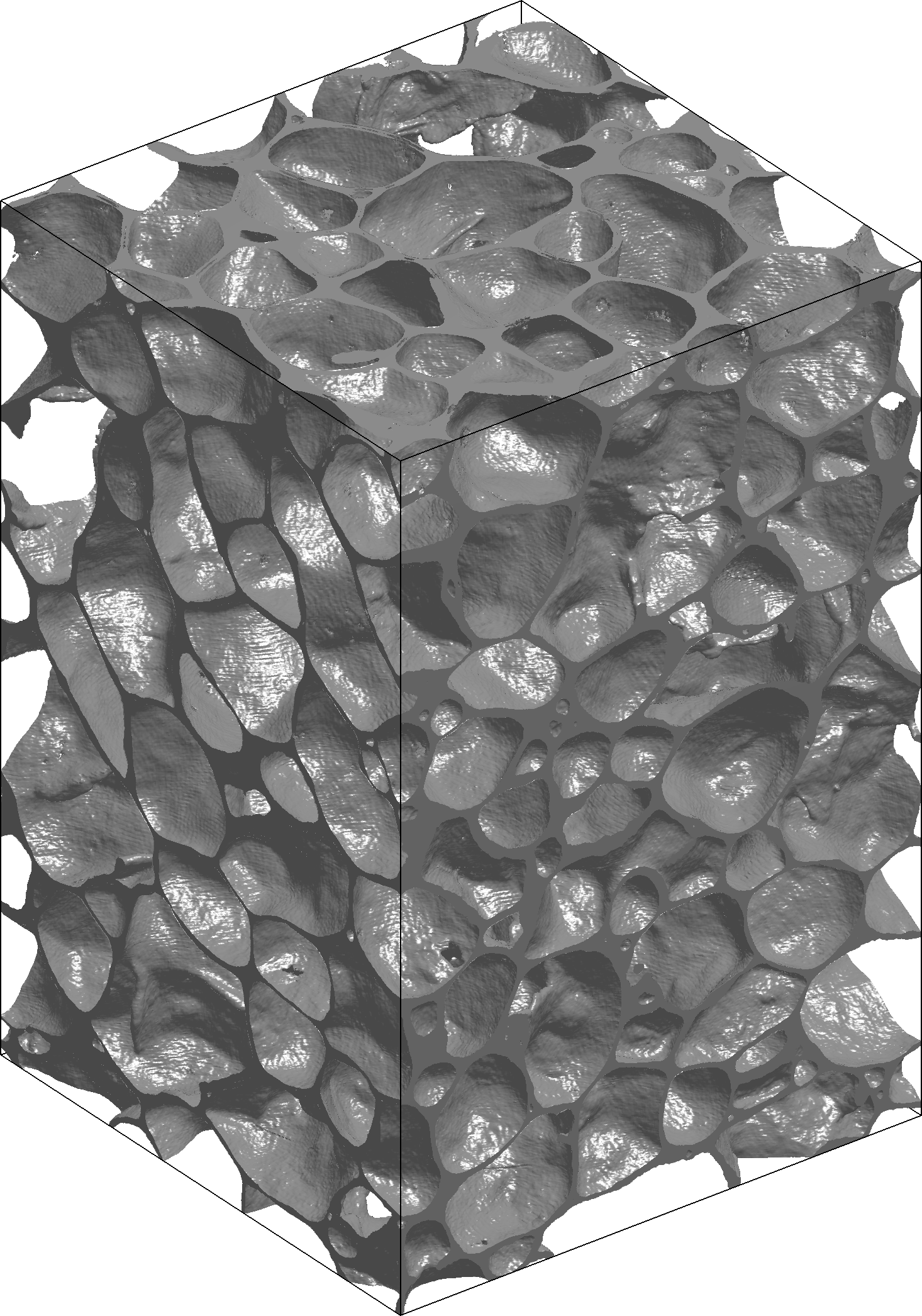}& 
    \includegraphics[height=2.85cm]{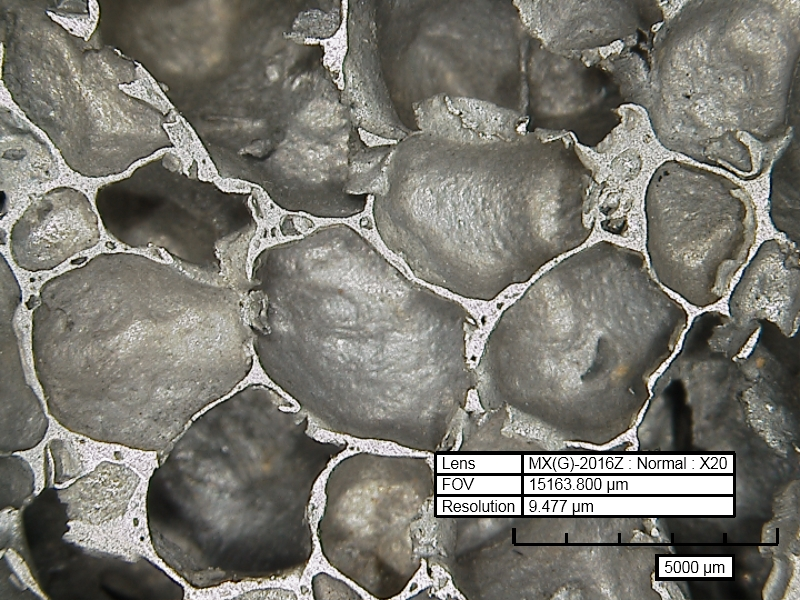}&
    \includegraphics[height=2.85cm]{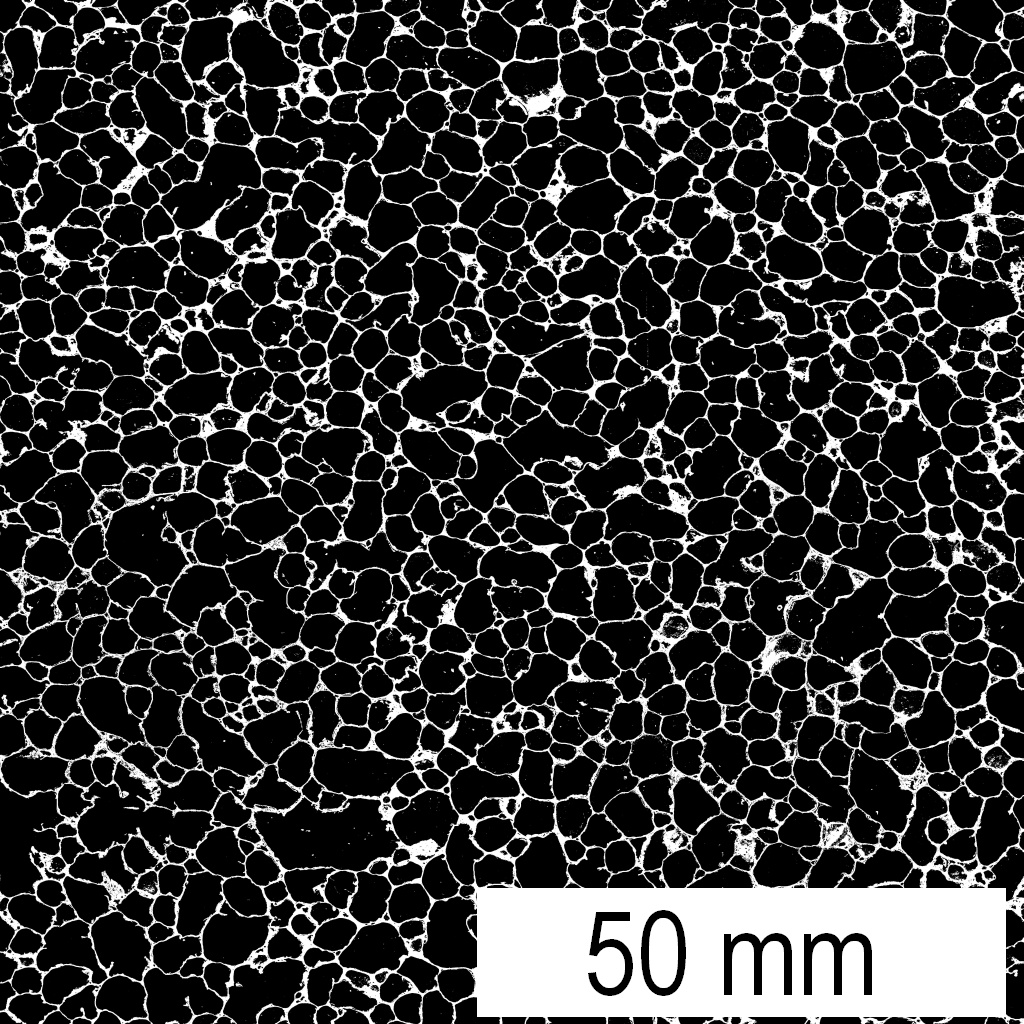}\\ 
	(a)&(b)&(c)
  \end{tabular}
  \caption{\alporas{} foam: a) Visualization of raw $\mu$-CT data, b) image from optical microscope, c) binary image of polished cross-section, courtesy of Ond\v{r}ej Jirou\v{s}ek~\cite{jirousek2012xray} (a) and Ji\v{r}\'{i} N\v{e}me\v{c}ek~\cite{nemecek2013twoscale} (b,c).}
  \label{fig:alporas}
\end{figure}

Two distinct phases, (i) the aluminum rich phase, and (ii) the phase of precipitates consisting of titanium and calcium, were identified by  N\v{e}me\v{c}ek\etal{} at the level of cell walls for the investigated specimen  displayed in~\Fref{fig:alporas}. In~\cite{nemecek2013twoscale} the Mori-Tanaka homogenization scheme was employed to estimate the values of effective Young's modulus of $70~\textrm{GPa}$ and Poisson's ratio of $0.35$ of the bulk material. In order to obtain comparable data with~\cite{nemecek2013twoscale}, the values were taken here as the reference parameters of the solid (ligament) phase for the homogenization at the level of foam. 

\section{Modeling strategy}
\label{sec:strategy}

Numerical homogenization is closely connected with the notion of Representative Volume Element (RVE), a characteristic sample of a microstructure large enough to provide us with the effective material properties. From this viewpoint, a method capable of rendering stochastic microstructure realizations of arbitrary sizes is desirable. 

\begin{figure*}[ht!]
	\centering
	
	\includegraphics[width=\textwidth]{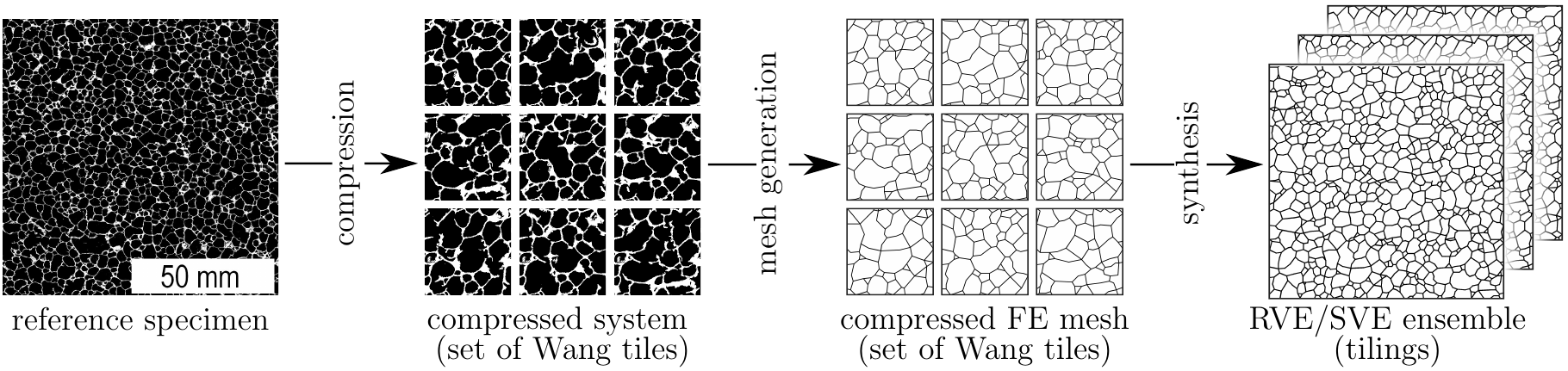}
	\caption{Illustration of adopted modeling strategy.}
	\label{fig:strategy_illustration}
\end{figure*}

Our approach rests on Wang tiling, a multicell generalization of the Periodic Unit Cell concept~\cite{novak2012compressing}, see~\Fref{fig:strategy_illustration}. We adopt the premise that a microstructure represented to a high degree of accuracy automatically provides us with a broad range of effective properties~\cite{niezgoda2010optimized,zeman2007random}. Starting with a binary two-dimensional image of a material, we compress the microstructure into a set of Wang tiles. Subsequently, the Finite Element (FE) mesh composed of planar beams conforming through the edges of identical codes is generated within each tile. Finally, in the numerical homogenization we benefit from the ability of the tiling concept to produce ensembles of microstructure realizations and corresponding computational models of arbitrary sizes using a stochastic tiling algorithm. 

\subsection{Wang tilings}
\label{sec:tiling_description}

\begin{figure}[ht!]
  \centering
  \setlength{\tabcolsep}{5pt}
  \begin{tabular}{ccc}
    \includegraphics[height=2cm]{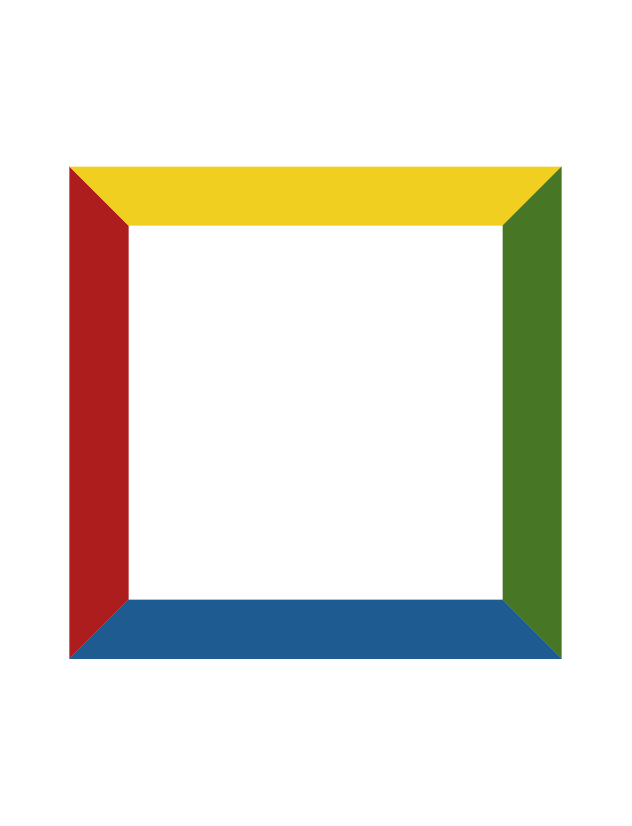} & \includegraphics[height=2.45cm]{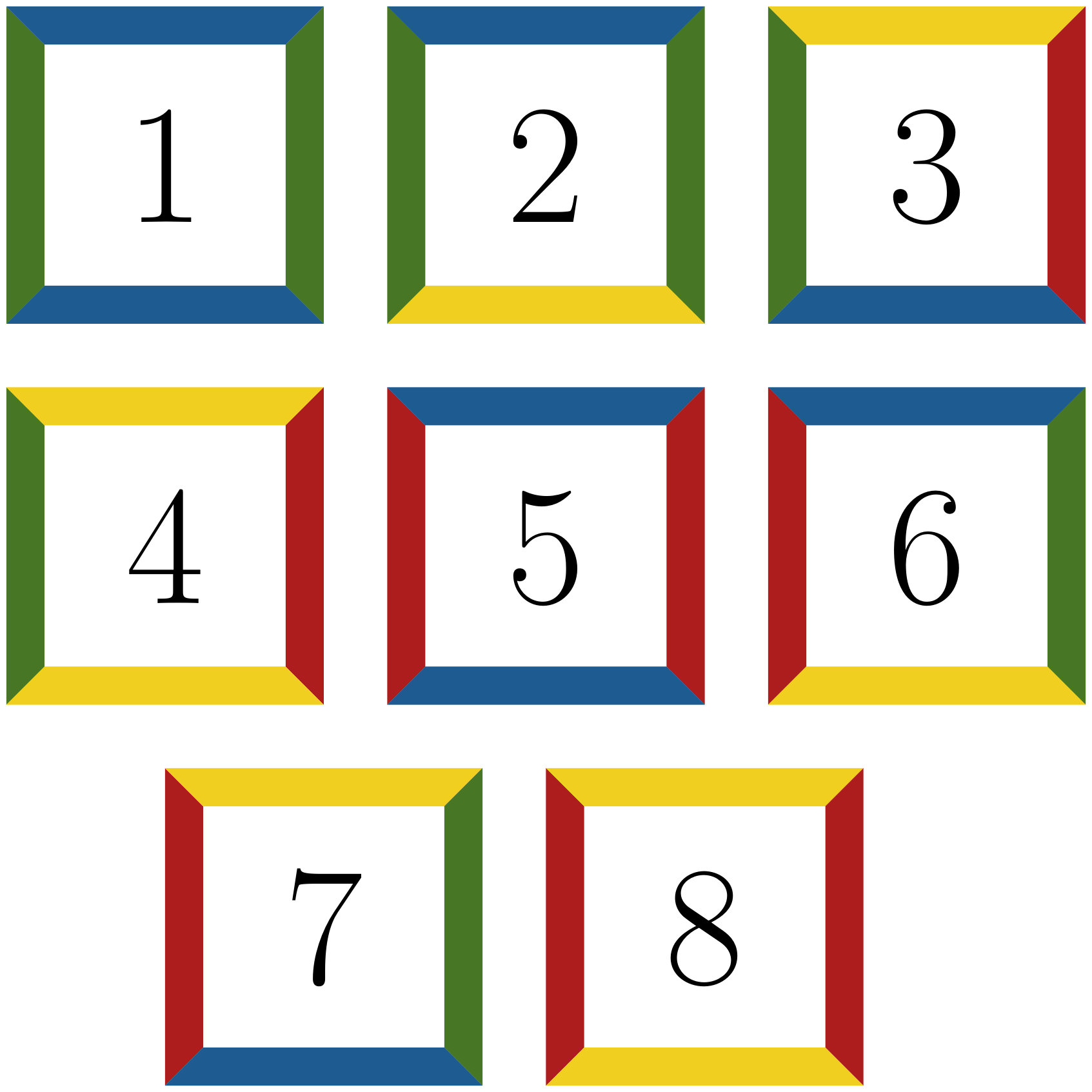} & \includegraphics[height=2.45cm]{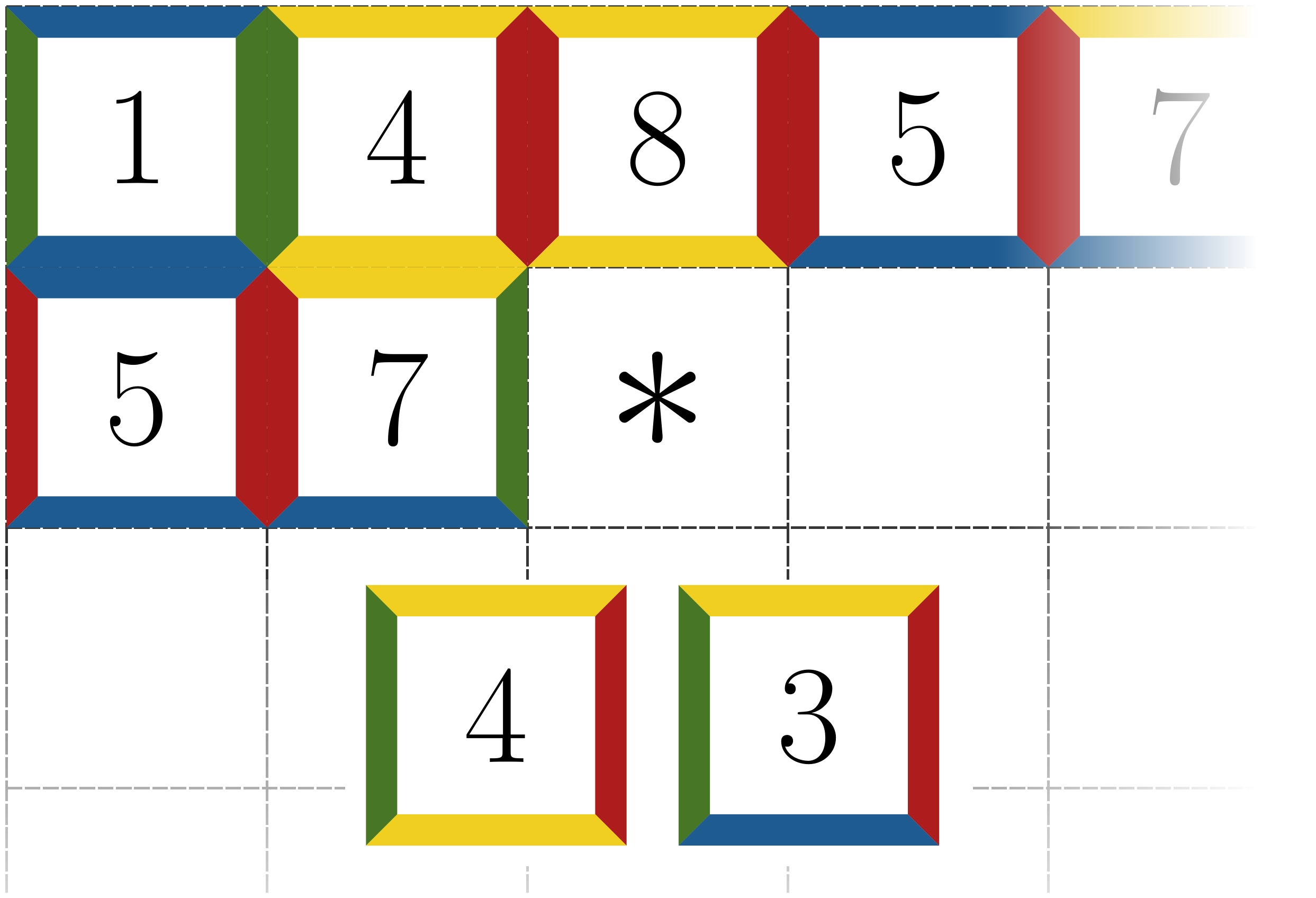} \\
    (a) & (b) & (c)
  \end{tabular}
  \caption{(Color online) Illustration of: a) Wang tile, b) set of 8 Wang tiles, c) single step of stochastic tiling algorithm.}
  \label{fig:tileillustration}
\end{figure}

Wang tiles are square domino-like pieces with codes assigned to their edges,~\Fref{fig:tileillustration}(a). The codes represent binding constrains such that only the tiles with the same codes on the adjacent edges may be placed next to each other when assembled into a tiling. By definition, the tiles can be neither rotated nor reflected~\cite{wang1961proving}, hence the two tiles of identical code sequences mutually rotated by $\frac{\pi}{2}$ are considered as different. The term tile set denotes the union of all tiles available for the tiling synthesis. The set is characterized with the number of tiles and the number of distinct codes on their horizontal and vertical edges. The composition of codes within the tile set governs the probability of occurrence of each tile in the tiling.

Specific families of tile sets can be identified. Each of the families usually comes with a unique tiling algorithm. A great attention, especially from Discrete Mathematics community, has been paid to the family of aperiodic sets in the last 50 years. However, in agreement with Computer Graphics applications~\cite{cohen2003wang}, we prefer stochastic sets to strictly aperiodic ones as the former provide us with more freedom in their design. The stochastic sets also correspond better with the purpose of modeling of materials with random/disordered microstructure. 

Tiles of a stochastic tile set can be easily assembled into tilings by making use of the stochastic tiling algorithm proposed by Cohen\etal{}~\cite{cohen2003wang}. First, a regular lattice of desired size is created, see~\Fref{fig:tileillustration}c. The grid is sequentially filled up with tiles in either column-by-column or row-by-row order. At each step, a tile is randomly chosen from a subset of tiles that satisfy edge constrains given by the previously placed tiles and positioned. The randomness is guaranteed by the existence of at least two distinct tiles for each admissible edge code combination on the top and left-hand side edges. A schematic of the procedure is displayed in \Fref{fig:tileillustration}(c). The tiles 4 and 3 from the tile set, \Fref{fig:tileillustration}(b), form the unique subset for one step in the stochastic tiling algorithm sketched in \Fref{fig:tileillustration}(c). Either of the tiles is randomly chosen and fixed at the spot indicated with the asterisk mark.

\subsection{Tile-based microstructure compression}

In order to make the Wang tile concept applicable to modeling of random heterogeneous materials, microstructural information has to be attributed to each tile such that a tiled domain (a tiling) corresponds to the original microstructure and the microstructure continuity is not violated across the congruent edges. By analogy to the Periodic Unit Cell (PUC) morphology design \cite{zeman2007random,yeong_reconstructing_1998}, a way to meet those requirements is to employ optimization methods, e.g. Simulated Annealing~\cite{novak2012compressing}. Although this approach is fully general, the computational cost is tremendous especially for larger sets, high microstructure resolution, multiphase media, and/or multi-point spatial descriptors. Therefore, in this study we employ the automatic design procedure proposed originally for compression of digital textures~\cite{cohen2003wang} and explored later in materials modeling \cite{doskar2015aperiodic}.

In brief, we take a number of samples from the reference specimen of the investigated microstructure. Each of the samples is associated with a unique edge code from the tile set. Each tile is then constructed as a square cut, rotated by $\frac{\pi}{4}$, out of partially overlapping samples that are positioned according to the tile edge codes they are associated with~\cite{cohen2003wang}. Since the cut goes diagonally through the samples, the microstructure continuity is automatically preserved as the sample leftovers are used as microstructural information for the tiles having the same codes on the opposite edges. The samples are seamlessly fused in the tile interiors by making use of the Image Quilting algorithm due to Efros\etal{}~\cite{efros2001image}. We have augmented the procedure by optimizing the design input parameters to achieve maximum proximity in spatial statistics of synthesized tilings and the original microstructure. Namely, volume fraction, two-point probability and two-point cluster functions were considered. This approach appears extremely efficient from the viewpoint of both the quality of compressed microstructures and computational overhead, see~\cite{doskar2015aperiodic} for details. Thus, after the compression based on the automatic tile design algorithm we have a binary representation of the Alporas microstructure in the form of a set of Wang tiles at hand.

\subsection{Discretization}

Validity of three-dimensional beam models for predicting elastic behavior of foams has been reported several times, e.g.~\cite{gibson1982mechanics,ashby2000metal,koudelka2012effective}. In the recent work~\cite{nemecek2013twoscale}, N\v{e}me\v{c}ek \etal{} presented promising results also for the two-dimensional beam representation of foam motivating our modeling strategy. In their approach the foam microstructure was represented as the Voronoi wired model consisting of straight beams with the centroids of individual pores taken as the Voronoi diagram seeds. The wired model was preferred to the two-dimensional finite element analysis as the latter would have required modifications to ligament walls thickness in order to fit the volume fraction from the weighting experiment. Its difference against the volume fraction determined from the area of ligaments in the reference planar scan was attributed to the preparation procedure, in which the saw cut was not perpendicular to most of the cell walls yielding the apparent solid phase volume be higher than the true one.

\begin{figure*}[ht!]
  \centering
  \setlength{\tabcolsep}{6pt}
  \begin{tabular}{cc}
    \includegraphics[width=0.48\textwidth]{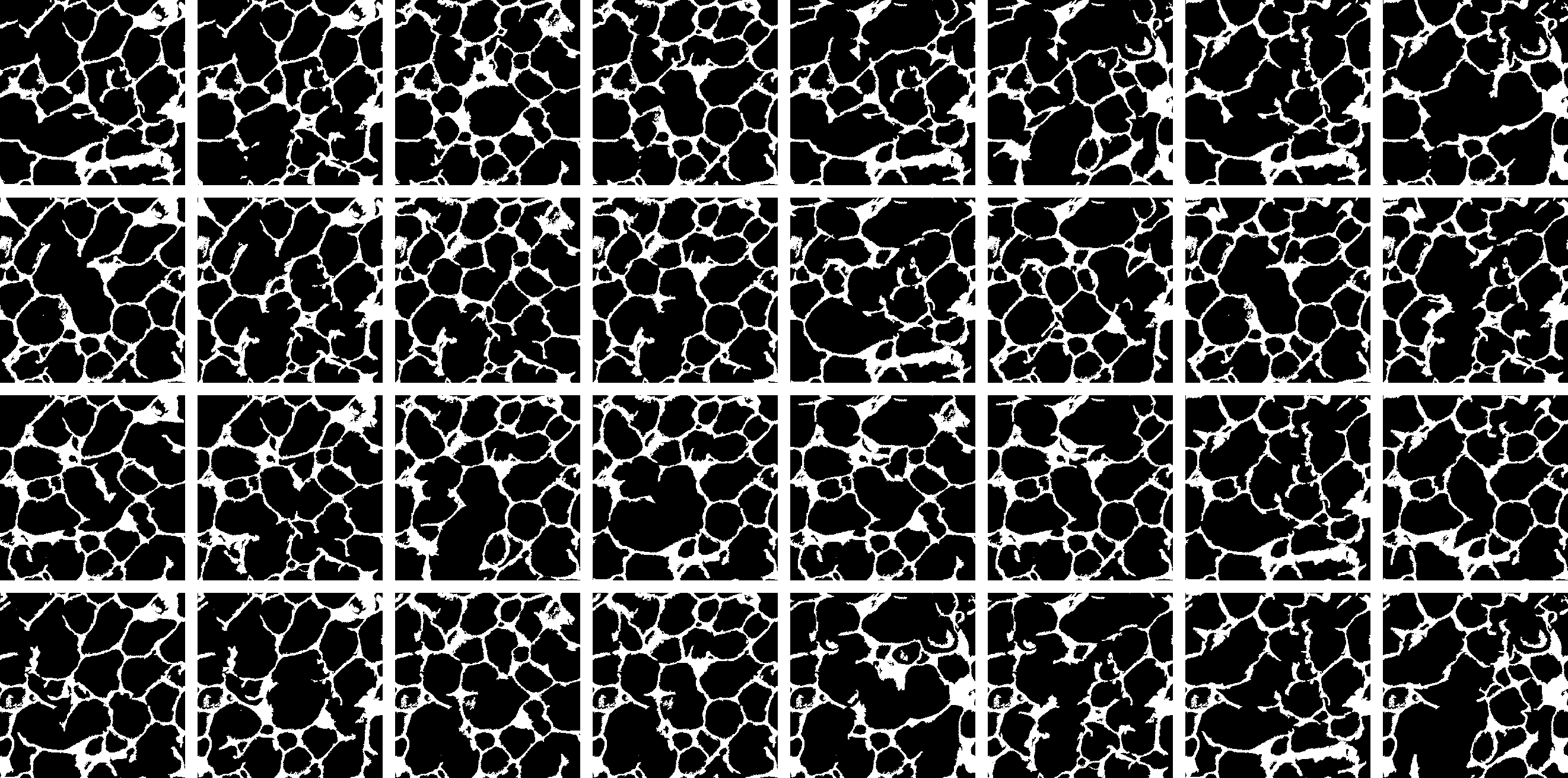} & \includegraphics[width=0.48\textwidth]{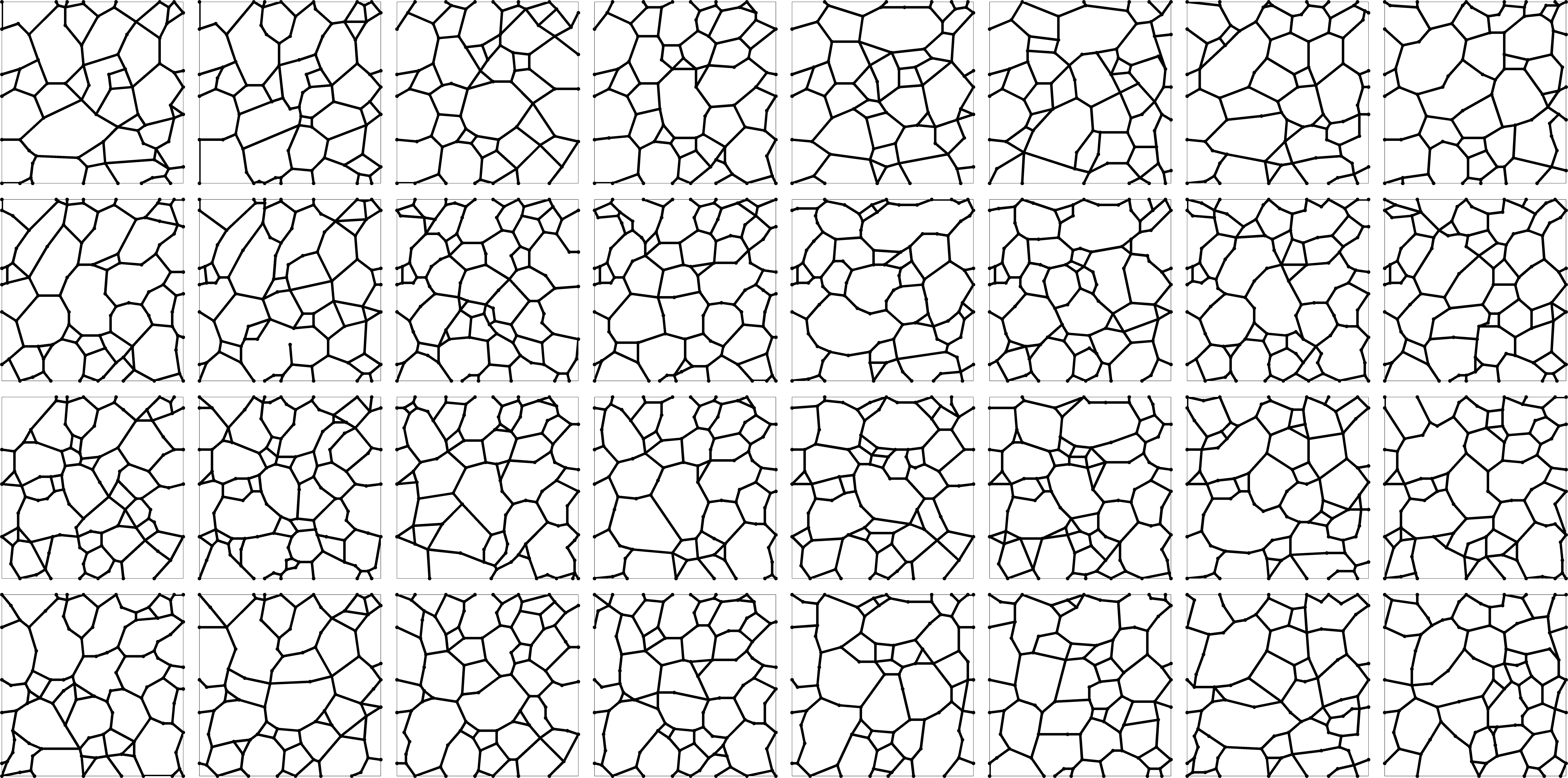}\\
    (a) & (b) 
  \end{tabular}
  \caption{\Alporas microstructure compressed within 32 tiles: (a) microstructure of individual tiles, (b) corresponding wired model (beam discretization of ligaments).}
  \label{fig:compression}
\end{figure*}

The size of the model used by authors of~\cite{nemecek2013twoscale} was limited by the extent of the given scanned image. However, this is not the case of Wang tiles, which enable to investigate series of models yielding adequate RVE dimensions. In addition, the beam geometry conforming with the real shape of ligaments allows us to explore the inaccuracy resulting from the Voronoi tessellation based idealization.
We started with the foam geometry compressed within the set of 32 tiles with four distinct edge codes on vertical and horizontal edges (8 in total), \Fref{fig:compression}a. First, a parametric beam discretization of tile geometries was created manually by employing a specifically designed GUI tool written in \Matlab{} environment\footnote{available at \url{http://mech.fsv.cvut.cz/~doskar/software/MeshMaker.zip}},~\Fref{fig:compression}b. It enabled us to control the mesh compatibility across tile edges and to fix ligaments disrupted by quilting or faults due to the sample preparation,~\Fref{fig:alporas}. 
Computational domains were assembled from the individual tiles by making use of the stochastic tiling algorithm described in~\Sref{sec:tiling_description}, the nodes on the coincident edges were identified, merged, and the entire record renumbered to maintain the sparsity of algebraic systems.

The cross-sectional parameters of individual beams were determined by analogy to~\cite{nemecek2013twoscale}. Beams were assumed straight and prismatic with a rectangular cross-section of the unit width. The total length of beams $L_\Sigma$ is given by the mesh geometry. Hence, the only degree of freedom, the depth of the cross-section $H$, is governed by the volume fraction of the solid phase $\rho$ and RVE volume/area $\measure{\domain}$. It reads as
\begin{equation}
  H= \frac{ \rho \measure{\domain} }{ L_\Sigma}
  \label{eq:h}
\end{equation}
Together with the assumption of unit beam width, \Eref{eq:h} further yields expressions for the cross-sectional area $A$ and the second moment of area $I$
\begin{equation}
	A = H, \quad I = \frac{1}{12} H^3
	\label{eg:crosssection_definition}
\end{equation}

\begin{figure*}[ht!]
  \centering
  \setlength{\tabcolsep}{2pt}
  \begin{tabular}{cccc}
    \includegraphics[width=0.24\textwidth]{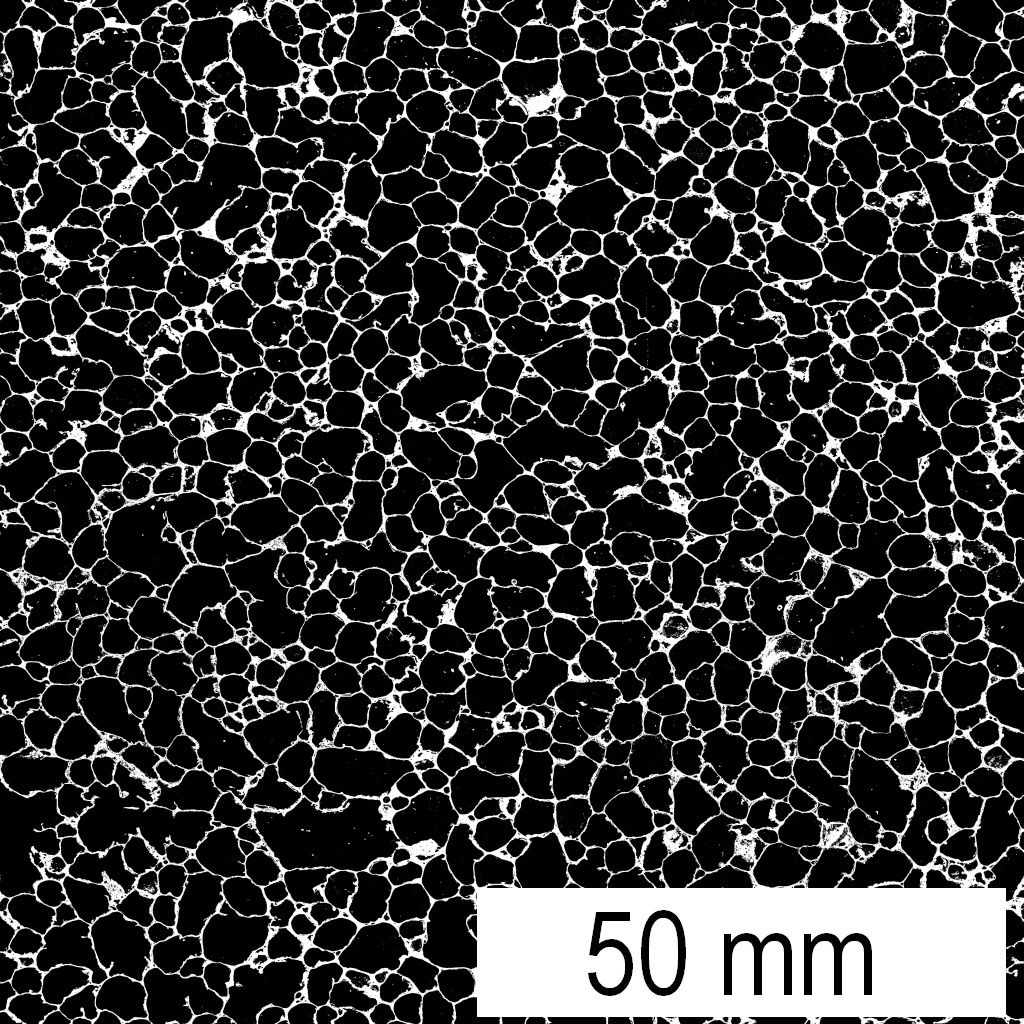} & \includegraphics[width=0.24\textwidth]{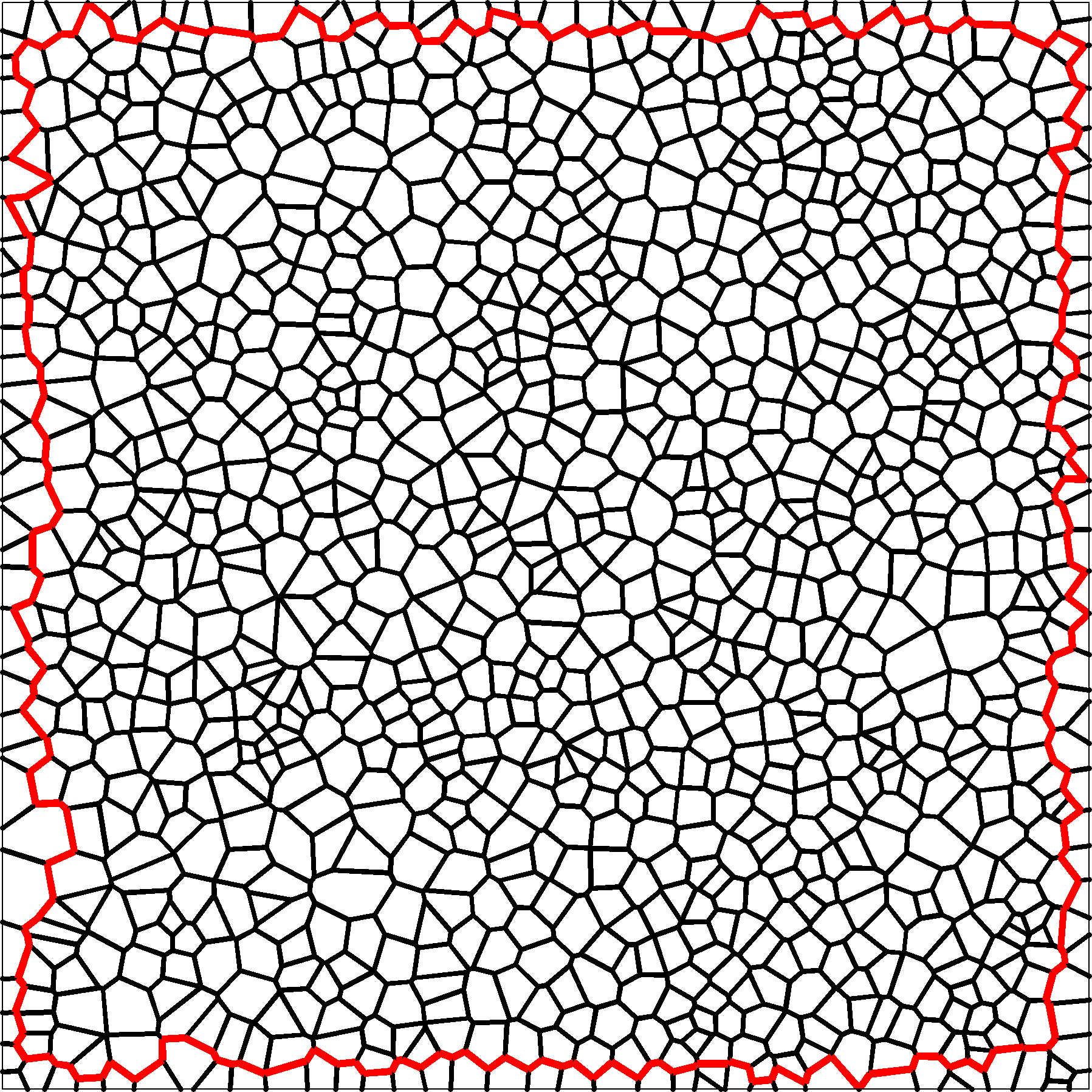} & \includegraphics[width=0.24\textwidth]{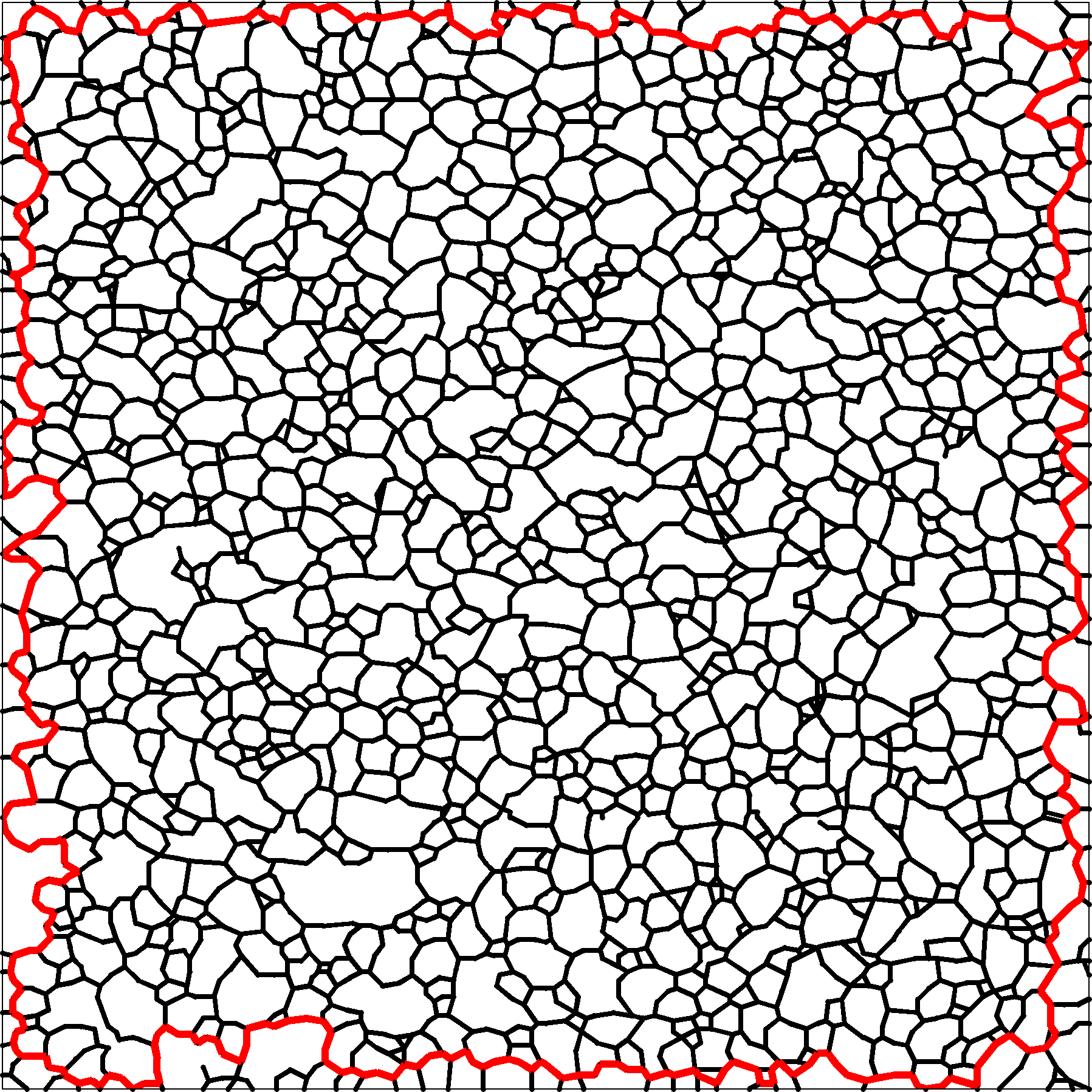} & \includegraphics[width=0.24\textwidth]{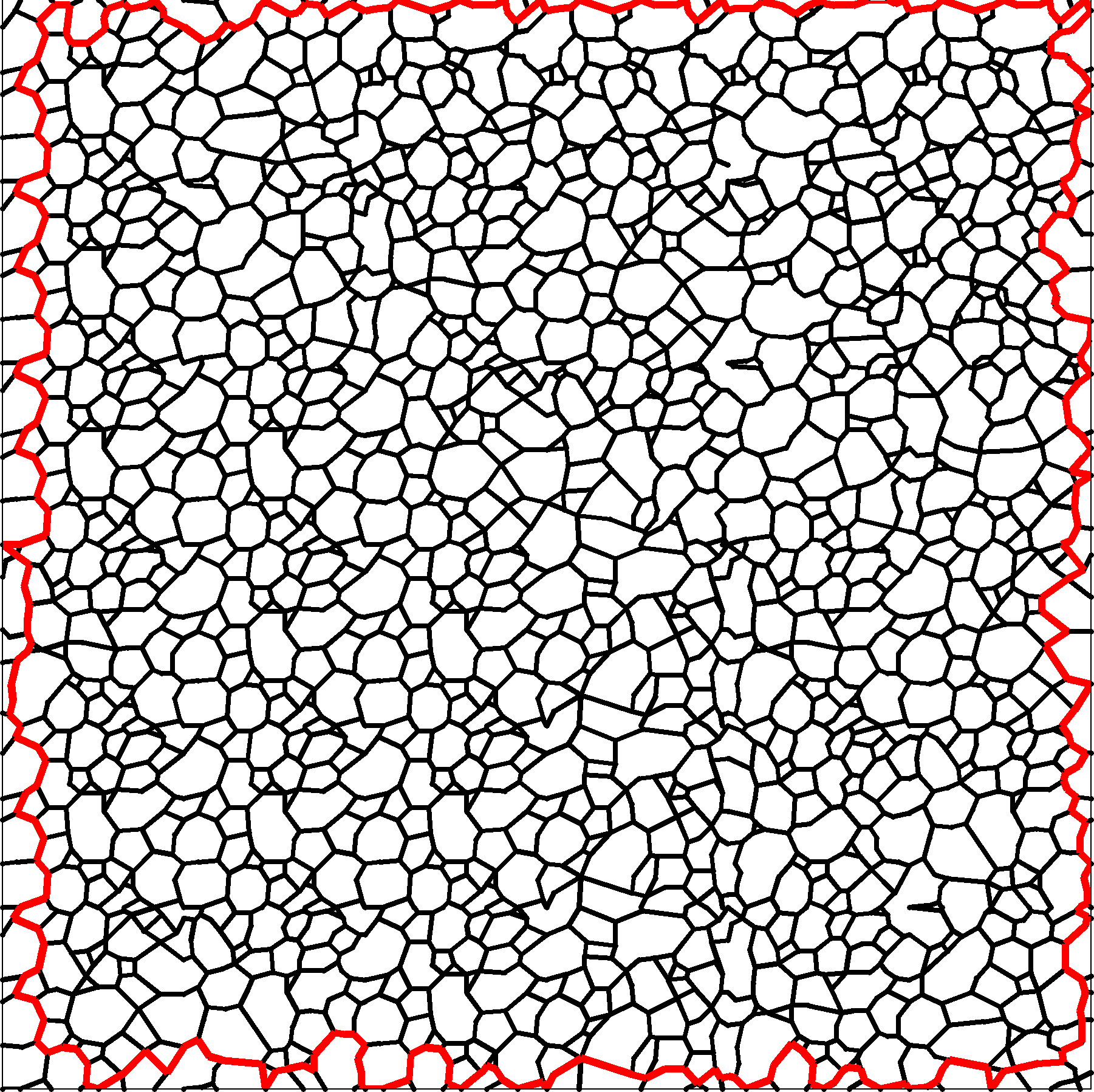} \\
    (a) & (b) & (c) & (d)     
  \end{tabular}
  \caption{(Color online) Comparison of various representations of \Alporas{} geometry, a) reference microstructure (courtesy of Ji\v{r}\'{i} N\v{e}me\v{c}ek~\cite{nemecek2013twoscale}), b) Voronoi mesh, c) mesh of real geometry, d) tiling-based representation of beam mesh ($6\times6$ tiling). The red line highlights boundary beams such that they coincide with $\boundary$.}
  \label{fig:meshcomparison}
\end{figure*}

In order to analyze precisely the effect of beam geometry, a separate beam model, \Fref{fig:meshcomparison}c, was created directly from the reference geometry (referred to as real geometry mesh onwards) in our in-house developed GUI tool. It was accompanied also with the microstructure represented as the Voronoi diagram, \Fref{fig:meshcomparison}b, to compare with the results published in~\cite{nemecek2013twoscale}. For visual comparison, a tiling-based geometry of similar size (composed of $6\times6$ tiles) is shown in \Fref{fig:meshcomparison}d. 

\section{Numerical homogenization}
\label{sec:homogenization}

In general, the homogenization process provides a homogeneous substitute for a heterogeneous material composed of constituents of different properties. Limiting the exposition to linear elasticity, with Hooke's Law 
\begin{equation}
	\tenss{\sigma}(\ex) = \tensf{\D}(\ex) \dcontr \tenss{\strain}(\ex) \,, \quad \textrm{for} \, \ex \in \domain \,,
	\label{eq:heterogeneousHooke}
\end{equation}
relating local stresses $\tenss{\sigma}(\ex)$ to strains $\tenss{\strain}(\ex)$ and $\domain$ being the microscale domain of interest representing a macroscopic material point, we wish to replace the heterogeneous material stiffness tensor $\tensf{\D}(\ex)$ with its homogeneous counterpart $\Dhom$ independent of microscale coordinates $\ex$. An equation analogical to~\Eref{eq:heterogeneousHooke}, namely,
\begin{equation}
	\tenss{\Sigma} = \Dhom \dcontr \tenss{\Strain} \,,
	\label{eq:homogeneousHooke}
\end{equation}
then relates the macroscopic strains $\tenss{\Strain}$ and stresses $\tenss{\Stress}$, defined as spatial averages $\avgs{\cdot}$
\begin{equation}
	\begin{split}
		\tenss{\Strain} &= \avgs{\tenss{\strain}(\ex)} = \frac{1}{ \measure{\domain} } \int_{\domain} \tenss{\strain}(\ex) \de{ \ex } \quad\textrm{and}\\ 
		\tenss{\Stress} &= \avgs{\tenss{\stress}(\ex)} = \frac{1}{ \measure{\domain} } \int_{\domain} \tenss{\stress}(\ex) \de{ \ex } \,,
	\end{split}
	\label{eq:average}
\end{equation}
respectively. 

Homogenization of parameters describing linear problems is a well resolved area nowadays. For simple microstructural compositions and relatively low contrast in constituent properties analytical homogenization schemes yield accurate estimates, often based on the Eshelby solution, or narrow bounds, e.q. Voigt-Reuss-Hill or Hashin-Shtrikman, of the sought homogenized property~\cite[and references therein]{sejnoha_micromechanics_2012}.
On the contrary, for materials with high contrast in properties of its constituents the actual microstructural geometry, simplified in the analytical homogenization methods, influences significantly the material response. Consequently, the analytical methods are suitable only for quick precursory estimates of the overall material stiffness. In the particular example of highly porous foams, even the closest estimates obtained with the Mori-Tanaka and the differential schemes differ from the experimental data by 20~\%~\cite{sevostianov_elastic_2006,sejnoha_micromechanics_2012}.

Hence, to determine the overall properties we resort to a numerical homogenization that, albeit more computationally intensive, gives us an insight into the size of the related \RVE{} and mechanics of deformation of the investigated microstructure. In particular, we adopt the first-order numerical homogenization method described comprehensively in work of Michel\etal{}~\cite{michel1999effective}. First, the general concept is recalled in the tensorial notation and the variational setting. Then, we reformulate the approach for a two-dimensional beam model, using the matrix notation. 
Numerical homogenization, unlike the analytical schemes, calls for a suitable computational domain -- \RVE{}, and for specific boundary conditions in order to render bounds of the effective property. We address these issues in~\Sref{sec:boundary_conditions} and discuss incorporation of different boundary conditions by means of the Lagrange multipliers.


\subsection{First order homogenization}
\label{sec:first_order_homogenization}

In the first order homogenization, the total displacement field $\tens{u}\atx$ can be decomposed as~\cite{michel1999effective}
\begin{equation}
  \tens{u}\atx= \tens{u}^{\tenss{\Strain}}\atx + \tens{u}^{\ast}\atx =  \tenss{\Strain} \scontr \ex + \tens{u}^{\ast}\atx \,, \quad \textrm{for} \, \ex \in \domain \,,
  \label{eq:udecomposition}
\end{equation}
where $\tens{u}^{\tenss{\Strain}}\atx$ is coupled with the homogeneous strains $\tenss{\Strain}$ and $\tens{u}^*\atx$ are displacement fluctuations induced by presence of material heterogeneity. Applying the symmetric gradient operator $\tens{\nabla}^s = \frac{1}{2} \left( \tens{\nabla} + \tens{\nabla}^\trn \right)$ to the displacement fields in \Eref{eq:udecomposition} provides us with the total strains in the form
\begin{equation}
  \tenss{\strain}\atx= \tens{\nabla}^s \tens{u}\atx = \tenss{E} + \tens{\nabla}^s \tens{u}^{*}\atx \,, \quad \textrm{for} \, \ex \in \domain \,.
  \label{eq:straindefinition}
\end{equation}
Thus, the strain energy density on microscale reads
\begin{equation}
\begin{split}
	\microdensity\left( \tens{x}, \tenss{\Strain}, \tens{\nabla}^s \tens{u}^{\ast}\atx \right) =\\  \half \left( \tenss{\Strain} + \tens{\nabla}^s \tens{u}^{*}\atx \right) \dcontr \tensf{D}\atx \!\dcontr\! \left( \tenss{\Strain} + \tens{\nabla}^s \tens{u}^{*}\atx \right) .
	\label{eq:microdensity}
\end{split}
\end{equation}

For the purpose of  variational formulation, we distinguish between the true fields $\tenss{\Strain}$ and $\tens{u}^{*}$ and the corresponding test fields denoted with tilde. 
Let $\mathcal{C}$ be a set of all admissible fluctuation displacement fields complying with the boundary conditions, further specified in~\ref{sec:boundary_conditions}. For arbitrary test fields $\tilde{\tens{u}}^{*} \in \mathcal{C}$ and $\tilde{\tenss{\Strain}} \in \setR^{d\times d}_{\textrm{sym}}$, where $d$ is the dimension of the model under consideration ($d = 2$ in our particular case), we define the macroscopic strain energy density $\macrodensity$ as
\begin{equation}
	\macrodensity ( \tilde{\tenss{\Strain}}, \tilde{\tens{u}}^{*}  ) = \avgs{  \microdensity ( \tens{x}, \tilde{\tenss{\Strain}}, \tens{\nabla}^s \tilde{\tens{u}}^{\ast}\atx  ) } \,.
	\label{eq:variational_macrodensity}
\end{equation}
Assume the stress $\tenss{\Stress}$ to be prescribed on macroscale. The elastic energy potential $\potential(\tilde{\tenss{\Strain}},\tilde{\tens{u}}^{*})$ for a macroscopic material point is then specified in the form
\begin{equation}
	\potential(\tilde{\tenss{\Strain}},\tilde{\tens{u}}^{*}) = \macrodensity (\tilde{\tenss{\Strain}},\tilde{\tens{u}}^{*} ) - \tenss{\Stress} \dcontr \tilde{\tenss{\Strain}} \,.
	\label{eq:energy_potential}
\end{equation}

The true fields $\tenss{\Strain}$ and $\tens{u}^{*}$ are obtained by minimizing the latter equation following the minimum total potential energy principle. Finally, combining the stationary conditions of \Eref{eq:energy_potential} and \Eref{eq:homogeneousHooke} results in the expression
\begin{equation}
	\Dhom = \frac{\partial^2 \macrodensity}{\partial {\tilde{\tenss{\Strain}}}^2} \left( \tenss{\Strain}, \tens{u}^{*} \right) \,.
	\label{eq:Dhom}
\end{equation}


\subsection{Discretization - Macroscopic Degrees of Freedom}
\label{sec:macroscopic_degrees}

As discussed in the previous sections, we assume $\domain$ to be a planar model discretized by straight beams. The model is embedded in the $x\mhyphen{}z$ plane, thus the unknowns of the beam model are nodal displacements $u$, $w$ along $x$, $z$ axes, and rotations $\phi$ about $y$. The dual quantities are normal forces $N$, shear forces $V$ and bending moments $M$ acting at beam ends (denoted by subscript indices $1$ and $2$). It holds
\begin{equation}
	\sevek{f}_e= \semtrx{K}_e \sevek{u}_e \,,
	\label{eq:local}
\end{equation}
where $\sevek{f}_e = \{ N_1, V_1, M_1, N_2, V_2, M_2 \}^\trn$, $\semtrx{K}_e$ is the element stiffness matrix calculated according to the Timoshenko theory~\cite{timoshenko1951theory}, and $\sevek{u}_e = \{ u_1, w_1, \phi_1, u_2, w_2, \phi_2 \}^\trn$.

The displacement decomposition in \Eref{eq:udecomposition}, rewritten in the matrix form, gives
\begin{equation}
  \sevek{u}_e = \begin{bmatrix} \sevek{I} & \sevek{A}_e \end{bmatrix}  \begin{Bmatrix} \sevek{u}_e^* \\ \sevek{\Strain} \end{Bmatrix} \,,
  \label{eq:udecompositionmtrx}
\end{equation}
where $\sevek{I}$ is the identity matrix of $6\times6$ entries and
\begin{equation}
\arraycolsep=2.5pt
  \semtrx{A}^\trn_e = 
  \begin{bmatrix}  
    x_1 &0 &0 &x_2 &0 &0\\
    0 &z_1 &0 &0 &z_2 &0\\
    \frac{1}{2}z_1 &\frac{1}{2}x_1 &0 &\frac{1}{2}z_2 &\frac{1}{2}x_2 &0
  \end{bmatrix}
\end{equation}
couples the nodal degrees of freedom (DOFs) with the applied macroscopic strains $\sevek{\Strain}= \{ \Strain_{x}, \Strain_{z}, \mathit{\Gamma_{xz}} \}^\trn$. Thus, the components of the macroscopic strain play the role of additional macroscopic degrees of freedom of the discretized system, cf.~\cite{michel1999effective}.

In the variational setting, contributions of individual finite elements to the overall strain energy density are
\begin{equation}
\begin{split}
	\macrodensity_e (\tilde{\sevek{\Strain}},\tilde{\sevek{u}}^*) &= \frac{1}{2\measure{\domain}} \tilde{\sevek{u}}^\trn_e \semtrx{K}_e \tilde{\sevek{u}}_e \\&=  \frac{1}{2\measure{\domain}} \begin{Bmatrix} \tilde{\sevek{u}}_e^* \\ \tilde{\sevek{\Strain}} \end{Bmatrix}^\trn  \underbrace{ \begin{bmatrix} \sevek{I} & \sevek{A}_e \end{bmatrix}^\trn \semtrx{K}_e \begin{bmatrix} \sevek{I} & \sevek{A}_e \end{bmatrix} }_{ \semtrx{K}_e^{\mathrm{ex}} } \begin{Bmatrix} \tilde{\sevek{u}}_e^* \\ \tilde{\sevek{\Strain}} \end{Bmatrix} \,,
	\label{eq:mtrxenergylox}
\end{split}
\end{equation}
where $\semtrx{K}_e^{\mathrm{ex}}$ is the extended element stiffness matrix of the form
\begin{equation}
 \semtrx{K}^\mathrm{ex}_e =
  \begin{bmatrix} 
    \semtrx{K}_e &  
    \semtrx{K}_e \semtrx{A}_e\\
    \semtrx{A}_e^\trn \semtrx{K}_e& 
    \semtrx{A}_e^\trn  \semtrx{K}_e \semtrx{A}_e
  \end{bmatrix} \,.
  \label{eq:Ke_ex}
\end{equation}
With the extended discrete system (\ref{eq:mtrxenergylox}) at hand, the macroscopic strain energy can be written as
\begin{equation}
	\macrodensity(\tilde{\sevek{\Strain}},\tilde{\sevek{u}}^*) = \frac{1}{2\measure{\domain}} \begin{Bmatrix} \tilde{\sevek{u}}^* \\ \tilde{\sevek{\Strain}} \end{Bmatrix}^\trn \semtrx{K}^\mathrm{ex} \begin{Bmatrix} \tilde{\sevek{u}}^* \\ \tilde{\sevek{\Strain}} \end{Bmatrix} \,,
	\label{eq:discreteStrainEnergy}
\end{equation}
where $\tilde{\sevek{u}}^* = \assembly_{e=1}^{n_{e}}\tilde{\sevek{u}}^*_e$, $\semtrx{K}^\mathrm{ex}$ takes the form
\begin{equation}
\semtrx{K}^\mathrm{ex}=
\assembly_{e=1}^{n_{e}}\semtrx{K}^\mathrm{ex}_e =
\begin{bmatrix} 
\semtrx{K}^\mathrm{ex}_{11} &  
\semtrx{K}^\mathrm{ex}_{12} \\ 
\semtrx{K}^\mathrm{ex}_{21}  & 
\semtrx{K}^\mathrm{ex}_{22} 
\end{bmatrix} \,,
\label{eq:extendedstiffnessglob}
\end{equation}
and $\assembly_{e=1}^{n_{e}}$ is the finite element assembly operator. The submatrices $\semtrx{K}^\mathrm{ex}_{ij}$ arise from the assembly of the submatrices on the right-hand side of~\Eref{eq:Ke_ex} and $\nelem$ denotes the number of elements.

\subsection{RVE size and boundary conditions}
\label{sec:boundary_conditions}

The above procedure yields the homogenized parameters providing the computational domain coincides with the \RVE{} related to the investigated microstructure and physical phenomenon. According to Hill~\cite{hill1963elastic}, \RVE{} is the sample of a material that incorporates enough geometrical information to render the effective stiffness parameters regardless the prescribed boundary conditions as long as they would result in the uniform stress and strain states if the microstructure were homogeneous. For smaller domains, here called Statistical Volume Elements (SVEs), \Eref{eq:Dhom} gives only the boundary condition dependent apparent stiffnesses. Thus, when homogenizing, the size of $\domain$ is usually increased step by step until the apparent properties resulting from different boundary conditions either coincide, i.e. meeting Hill's criterion~\cite{hill1963elastic}, or the variance in the sought quantity computed from different realizations of the same size reaches a given limit~\cite{kanit2003determination}.

Among the admissible boundary conditions yielding the uniform strain in the homogeneous medium, three are commonly used. The periodic boundary conditions show the fastest convergence as the size of SVEs approaches that of RVE, however they render only the estimate of the effective stiffness and are cumbersome to impose if the microstructure is in general non-periodic, which is also the case of \Alporas{}. The remaining two, Kinematic Uniform Boundary Conditions (KUBC) and Static Uniform Boundary Conditions (SUBC), are known to result in the upper and lower bounds of the apparent stiffness coefficients of each SVE, refining so the Voigt-Reuss-Hill bounds~\cite{kanit2003determination,huet1990application}. To the best of our knowledge, the exact form of SUBC is unknown for beam models. We thus circumvent this drawback prescribing the Minimal Kinematic Boundary Conditions (mKBC) due to Mesarovic and Padbidri~\cite{mesarovic2005minimal}, since these were shown to equal SUBC~\cite{miehe2003computational}. 

To preserve the structure of the resulting algebraic system, both boundary conditions are prescribed as an additional constraint\footnote{We assume that the assembly operator $\assembly_{e=1}^{n_{e}}$ does not involve boundary conditions.} in the form
\begin{equation}
	\semtrx{C} \tilde{\sevek{u}}^{*} = \sevek{0}
	\label{eq:boundary_constrain}
\end{equation}
defining so the kinematically admissible nodal displacements and rotations stored in $\tilde{\sevek{u}}^{*}$. The boundary $\boundary$ has been determined from the beams fully contained in the domain, highlighted in red (light gray in BW version) in \Fref{fig:meshcomparison}. 

In the case of KUBC, the displacement fluctuations have to vanish at the boundary, i.e. $\tilde{\tens{u}}^{*}\atx = 0$ for $\ex \in \boundary$ . In the discrete beam model we prescribe zero fluctuation displacements $u$ and $v$ for all nodes at the boundary, the corresponding rotations $\phi$ are left unknown. With $\nboundarynodes$ boundary nodes and $\nallnodes$ nodes in total, the matrix of constraints $\semtrx{C}^{\Ckubc}$ for the KUBC case consists of $ \left[ 2 \nboundarynodes \times 3 \nallnodes \right]$ components and is defined as
\begin{equation}
	C_{ij}^{\Ckubc} = 
			\left\{ \begin{array}{ll}
				1\,, & \mbox{for} \, i = j \in \mathcal{G} \,, \\
				0\,, & \mbox{otherwise,}
			\end{array} \right.
\end{equation}
where $\mathcal{G}$ is the set of all code numbers corresponding to the fixed fluctuation displacements.

In contrast to KUBC, mKBC~\cite{miehe2003computational} do not require the fluctuation field to vanish at the boundary point-wise but they enforce the field to vanish on average. The definition of the macroscopic strain tensor $\tenss{\Strain}$, \Eref{eq:average}, along with the local strain field decomposition, \Eref{eq:straindefinition}, yields the minimal condition posed on $\tilde{\tens{u}}^{*}$ (also called the normalization condition~\cite{fish2005multiscale}) in the form
\begin{equation}
	\tens{0} = \frac{1}{\measure{\domain}} \int_{\domain}{ \tens{\nabla}^s \tilde{\tens{u}}^{*} }\de{\ex} =  \frac{1}{\measure{\domain}} \int_{\boundary}{ \tens{n} \otimes^{s} \tilde{\tens{u}}^{*} }\de{\ex} \,,
	\label{eq:minimal_condition}
\end{equation}
where $\tens{n}$ is the outer normal and $\otimes^{s}$ denotes the symmetric part of the tensor product defined as $\left( \tens{n} \otimes^{s} \tilde{\tens{u}}^{*} \right) = \half \left( \tens{n} \otimes \tilde{\tens{u}}^{*} + \tilde{\tens{u}}^{*} \otimes \tens{n} \right)$. In addition to the three conditions defined by~\Eref{eq:minimal_condition} (for the two-dimensional setting), we prescribe zero fluctuations $u^*$ and $w^*$ at the bottom-left and $w^*$ at the bottom-right corners to restrain the rigid body modes. Plugging the shape functions of the beam element in~\Eref{eq:minimal_condition} along with the three additional conditions mentioned above give rise to the matrix of constraints $\semtrx{C}^{\Cminkbc}$ with $ \left[ (3 + 3) \times 3 \nboundarynodes \right] $ entries, see~\Aref{s:minimalKBC} for additional details.

Having the matrices $\semtrx{C}^{\Ckubc}$ and $\semtrx{C}^{\Cminkbc}$ at hand, the constrained minimization of \Eref{eq:energy_potential} can be reformulated by making use of the Lagrange multipliers $\tilde{\sevek{\lambda}}$, which act as reactions and generalized reactions for  KUBC and mKBC, respectively. To this goal, we define the energy potential $\bar{J}(\tilde{\sevek{u}}^{*},\tilde{\sevek{\Strain}},\tilde{\sevek{\lambda}})$ similarly to \Eref{eq:energy_potential}
\begin{equation}
\bar{J}(\tilde{\sevek{u}}^{*},\tilde{\sevek{\Strain}},\tilde{\sevek{\lambda}}) = \frac{1}{\measure{\domain}} \!\left(\! \half \begin{Bmatrix} \tilde{\sevek{u}}^{*} \\ \tilde{\sevek{\Strain}} \end{Bmatrix}^{\!\!\trn}\!\!\semtrx{K}^\mathrm{ex}\!\begin{Bmatrix} \tilde{\sevek{u}}^{*} \\ \tilde{\sevek{\Strain}} \end{Bmatrix} + \tilde{\sevek{\lambda}}^{\trn} \semtrx{C} \tilde{\sevek{u}}^{*}\!\right)\!\! - \sevek{\Stress}^\trn \tilde{\sevek{\Strain}} \,.
\label{eq:final_potential}
\end{equation}
The stationary conditions posed on $\bar{J}$, namely, 
\begin{equation}
	\frac{\partial}{\partial \tilde{\sevek{u}}^{*} } \left. \bar{J} \right|_{\tilde{\sevek{u}}^{*}=\sevek{u}^{*}}\!\!\!= 0\,,\quad
	\frac{\partial}{\partial \tilde{\sevek{\Strain}} } \left. \bar{J} \right|_{\tilde{\sevek{\Strain}}=\sevek{\Strain}}\!\!= 0\,,\quad
	\frac{\partial}{\partial \tilde{\sevek{\lambda}} } \left. \bar{J} \right|_{\tilde{\sevek{\lambda}}=\sevek{\lambda}}\!\!\!= 0\,,
\end{equation}
form the following algebraic system for the true fields
\begin{equation}
  \begin{Bmatrix} \sevek{0} \\ \sevek{\Stress} \\ \sevek{0} \end{Bmatrix} =  \frac{1}{\measure{\domain}} \begin{bmatrix} \semtrx{K}_{11}^\mathrm{ex} & \semtrx{K}_{12}^\mathrm{ex} & \semtrx{C}^\trn \\ \semtrx{K}_{21}^\mathrm{ex} & \semtrx{K}_{22}^\mathrm{ex} & 0 \\ \semtrx{C} & 0 & 0 \end{bmatrix} \begin{Bmatrix} \sevek{u}^{*} \\ \sevek{\Strain} \\ \sevek{\lambda} \end{Bmatrix} \,.
	\label{eq:constrained_system}
\end{equation}
Finally, condensing $\sevek{u}^{*}$ and $\sevek{\lambda}$ from~\Eref{eq:constrained_system} yields the homogenized stiffness matrix $\Dhommtrx$ equivalently to~\Eref{eq:Dhom}. For large and sparse systems of linear equations direct condensation of~\Eref{eq:constrained_system} would be inefficient. We thus compute individual columns of $\Dhommtrx$ in three consecutive steps as the responses $\sevek{\Stress}^{(i)}$  to the unit load cases $\sevek{\Strain}^{(i)}$ defined as $\sevek{\Strain} ^{(1)} = \begin{Bmatrix} 1, 0, 0\end{Bmatrix}^\trn$, $\sevek{\Strain} ^{(2)} = \begin{Bmatrix} 0, 1, 0\end{Bmatrix}^\trn$, and $\sevek{\Strain} ^{(3)} = \begin{Bmatrix} 0, 0, 1\end{Bmatrix}^\trn$.


\subsection{Extraction of isotropic parameters}

The approach outlined above leads to the homogenized material stiffness matrix with components encoding the anisotropic behavior. If the degree of anisotropy is small, it is convenient to describe the constitutive behavior of the homogenized material with scalar quantities $K$ and $G$, see \Tref{tab:alporasproperties}. To this goal, we perform spectral analysis of $\Dhommtrx$ to determine the effective moduli from its eigenvalues~\cite{doskar2013masterthesis}. For the plane strain conditions, the stiffness matrix of isotropic material $\semtrx{\D}^{\textrm{iso}}$ and its eigenvalues $\sevek{\lambda}^{\textrm{iso}}$ in terms of $K$ and $G$ read
\begin{equation}
\arraycolsep=2.9pt
  \semtrx{\D}^{\textrm{iso}}\!\!= 
  \begin{bmatrix}
    K + \frac{4}{3} G & K - \frac{2}{3}G & 0 \\
    K - \frac{2}{3} G & K + \frac{4}{3}G & 0 \\
    0 & 0 & G 
  \end{bmatrix}\!\!,\,
  \sevek{\lambda}^{\textrm{iso}}\!\!= 
  \begin{Bmatrix}
    G \\ 2 G \\ \frac{2}{3} G + 2 K
  \end{Bmatrix}
  \label{eq:DhomKG}
\end{equation}
Note, that the $\min\to\max$ ordering of the $\sevek{\lambda}^{\textrm{iso}}$ components holds for non-auxetic materials only, i.e. for those with $\nu > 0$. Since the calculated overall stiffness matrix $\Dhommtrx$ and the matrix eigenvalues $\sevek{\lambda}^{\homg}$ may not obey \Eref{eq:DhomKG} exactly we employ the Least Square Method to obtain the moduli as
\begin{equation}
	\begin{Bmatrix} G^{\homg} \\ K^{\homg} \end{Bmatrix} =  \begin{bmatrix} \frac{1}{5} & \frac{2}{5} & 0 \\ -\frac{2}{30} & -\frac{4}{30} & \frac{1}{2} \end{bmatrix} \sevek{\lambda}^{\homg} \,,
	\label{eq:leastsquareKG}
\end{equation}
see \Aref{s:leastsquare} for additional details.

\section{Results}
\label{sec:results}

The cross-sectional characteristics of beams, \Tref{tab:foamsections}, for the three different geometries, \Fref{fig:meshcomparison}, were derived through the approach outlined in~\Sref{sec:strategy} with the target ligament volume fraction set to $8.6~\%$ \cite{nemecek2013twoscale}.Note that the depth of beams (corresponding to the cross-section area $A$, recall \Eref{eg:crosssection_definition}) in the tiling-based representation is about $10~\%$ smaller than for the previous two. We attribute this reduction to the automatic tile design, namely local modifications to the geometry within the overlap region, and corrections induced by manual mesh generation which may resulted in longer beams and the corresponding smaller depth $H$.
\begin{table}[ht!]
  \caption{Cross-sectional characteristics of beams conforming with weighted volume fraction.}
  \label{tab:foamsections}
  \centering
  \renewcommand{\arraystretch}{1.35}
  \begin{tabular}{ccccc}
    \hline
    & & Voronoi & real-geometry & tiling-based\\\hline
    $A$ & [$\textrm{m}^2$] & $1.431\times10^{-04}$ & $1.469\times10^{-04}$ & $1.317\times10^{-04}$ \\
    $I$ & [$\textrm{m}^4$] & $2.443\times10^{-13}$  & $2.643\times10^{-13}$ & $1.904\times10^{-13}$ \\\hline
  \end{tabular}
\end{table}

We have computed the apparent parameters for square computational domains/tilings of sizes ranging from $1\times1$ up to $300\times300$ tiles. For each size, we generated 32 different stochastic realizations so that the number of realizations equals the number of distinct tiles in the set. Thus, results for $1\times1$ tilings correspond to the averages over the individual tiles. 

\begin{figure*}[ht!]
  \centering
  \begin{tabular}{cc}
    \includegraphics[width=0.40\textwidth]{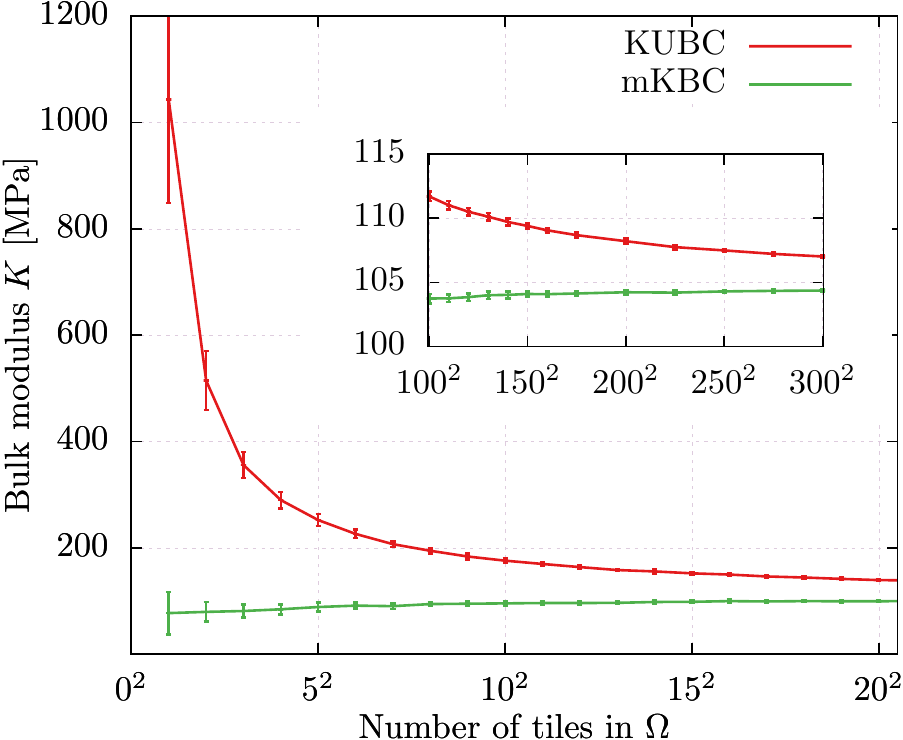} & \includegraphics[width=0.40\textwidth]{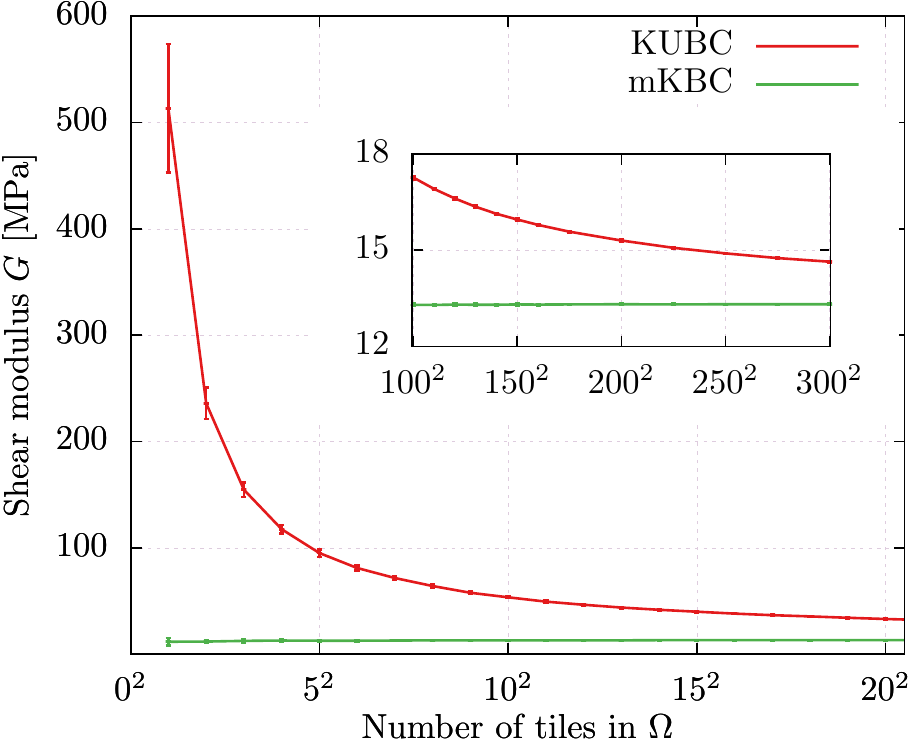} \\
    (a) & (b)\\
    \includegraphics[width=0.40\textwidth]{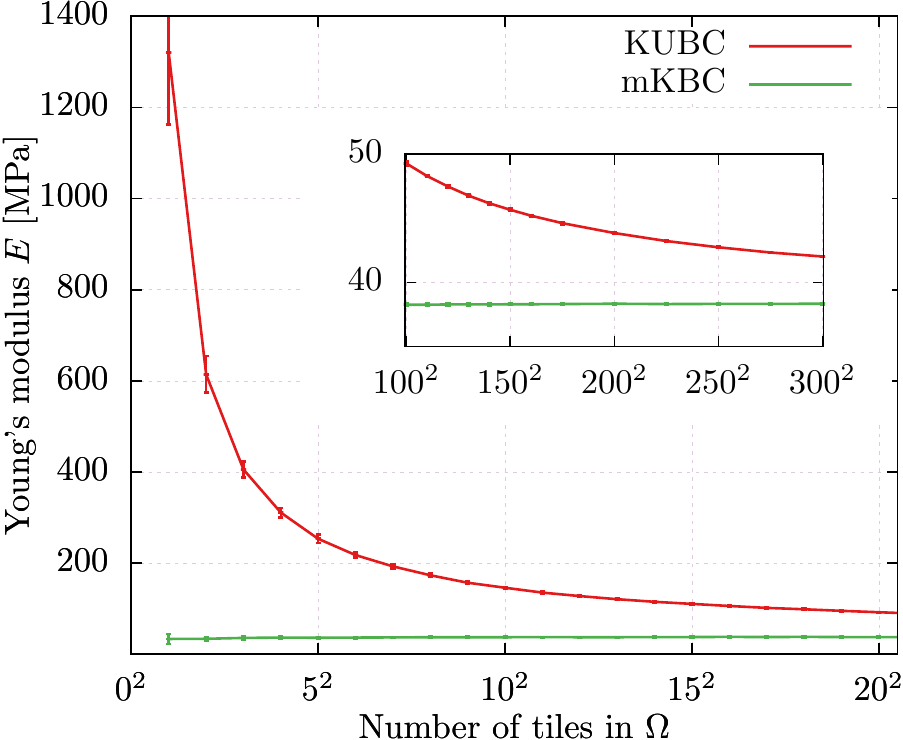} & \includegraphics[width=0.40\textwidth]{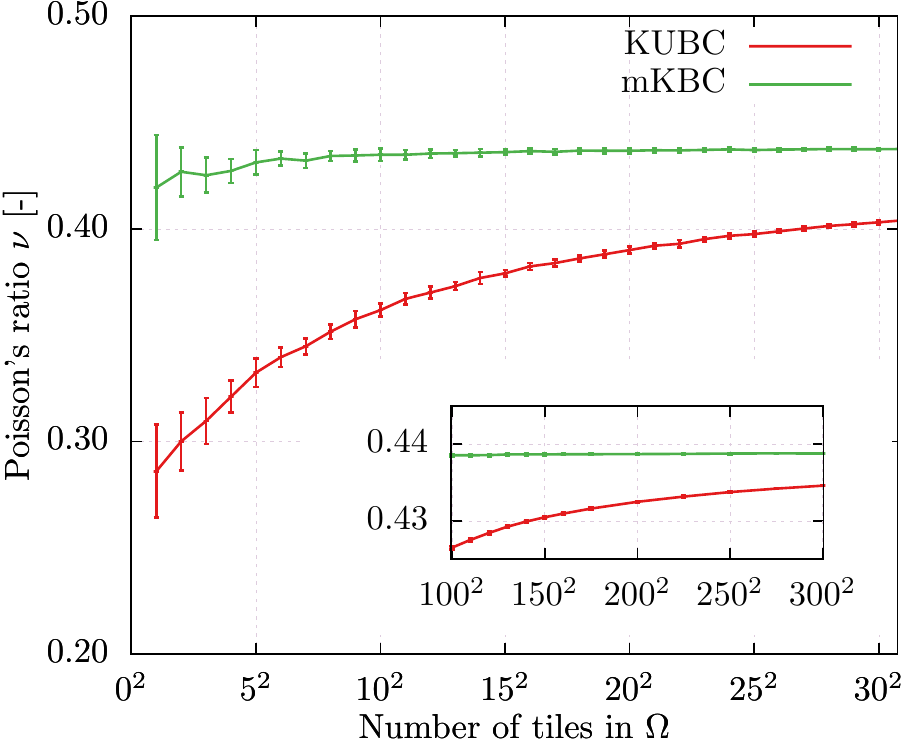} \\
    (c) & (d) 
  \end{tabular}
  \caption{(Color online) Relationship between homogenized elastic parameters and dimensions of computational domains: (a) Bulk modulus~$K$, (b) Young's modulus~$E$, (c) Shear modulus $G$, (d) Poisson's ratio $\nu$.}
  \label{fig:results}
\end{figure*}

\begin{figure}
	\centering
	\includegraphics[width=0.8\columnwidth]{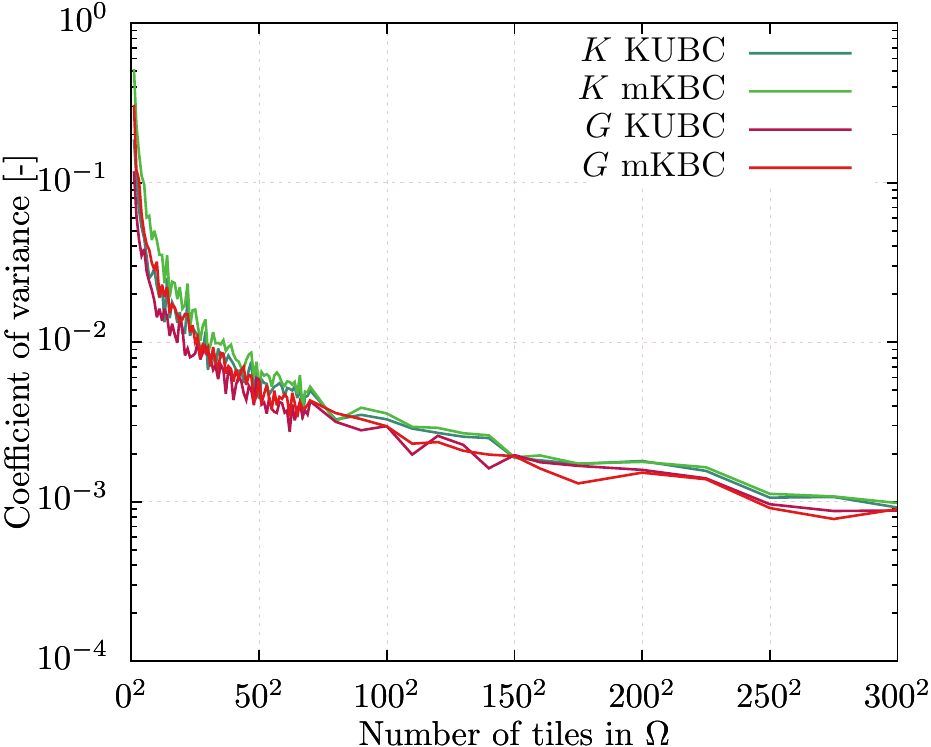}
	\caption{(Color online) Coefficient of variation of the upper and lower estimates on $K$ and $G$.}
	\label{fig:CoV}
\end{figure}

The means of the effective stiffness coefficients for models consisting of $300\times300$ tiles are enlisted in \Tref{tab:foamresults} together with the homogenized parameters arising from the Voronoi and real-geometry meshes as presented in~\Fref{fig:meshcomparison}.
\begin{table}[ht!]
	\centering
	\caption{Apparent elastic parameters of individual models.}
	\label{tab:foamresults}
	\setlength{\tabcolsep}{3pt}
	{{\footnotesize
	\renewcommand{\arraystretch}{1.25}
	\begin{tabular}{llcccc}
	\hline
	\multirow{2}{*}{} & & $K^{\homg}$ & $E^{\homg}$ & $G^{\homg}$ & $\nu^{\homg}$ \\
	& & [MPa] & [MPa] & [MPa] & [-] \\ \hline
	\multirow{2}{*}{Voronoi} & KUBC & 1435 & 238 & 81 & 0.472 \\
	& mKBC & 1027 & 51 & 17 & 0.492 \\ \hline
	\multirow{2}{*}{real-geometry} & KUBC & 203 & 220 & 84 & 0.319 \\
	& mKBC & 61 & 34 & 12 & 0.409 \\ \hline
	\multirow{2}{*}{tiling-based} & KUBC & 107(144) & 42(57) & 15(20) & 0.435(0.434) \\
	& mKBC & 104(141) & 38(53) & 13(18) & 0.439(0.438) \\\hline
	\end{tabular}
	}}
\end{table}
As expected, the upper bounds generated with KUBC and the lower bounds rendered by means of minKBC converge to each other with increasing size of computational domains, \Fref{fig:results}; even for tilings of $300\times300$ tiles they do not fully coincide. Nevertheless, we consider the difference between the bounds acceptable. This conclusion can be justified also in terms of the variance of the bounds. \Fref{fig:CoV} shows the asymptotic decay of coefficient of variation with the values less than $1~\permil$ in the vicinity of $300\times300$ tiling.

Since we failed to predict the reported stiffness parameters of \alporas{} foam, \Tref{tab:alporasproperties}, with the cross-sectional characteristics derived for tiling-based geometries, we performed the same computations also for cross-sectional parameters derived for the real-geometry mesh, \Tref{tab:foamsections}. The results, in brackets in \Tref{tab:foamresults}, were approximately $40~\%$ higher reflecting nearly linearly the $40\%$ increase in the second moment of area. However, they do not fall within the reported range of overall \alporas{} properties either.

\section{Conclusions and discussion}
\label{sec:conclusions}

We presented a modeling strategy for \alporas{} foam having at heart the synthesis of stochastic microstructure realizations based on the concept of Wang tiling. Besides the standard upper bounds on the effective stiffness given by Kinematic Uniform Boundary Conditions, we obtained the guaranteed lower bounds by means of minimal kinematic boundary conditions. With this formulation, shown to equal the common Static Uniform Boundary Conditions in~\cite{miehe2003computational}, we can avoid the question of an appropriate beam loading that would yield macroscopically uniform stress field. Hence, both bounds were rendered prescribing relevant boundary displacement.

Based on data in \Fref{fig:results}, we conclude that RVE for the two-dimensional model of highly porous materials should be about hundreds of the characteristic pore diameter length in dimensions. This finding corresponds with statements made in~\cite{niezgoda2010optimized} regarding the minimal size of RVE in the case of infinite contrast of phase properties. On the other hand, it contradicts the recommendation of Ashby\etal{}~\cite{ashby2000metal} who propose RVE size of approximately seven times the mean pore diameter. However, this recommendation is given for three-dimensional samples and does not need to be valid for planar analyses.

Comparison of the homogenized stiffness coefficients with the reference values reported by Ashby\etal{}~\cite{ashby2000metal}, Tabs. \ref{tab:alporasproperties} and \ref{tab:foamresults}, leads us to unambiguous conclusions. Despite the several times reported aptness of the \emph{spatial} wired model \cite{ashby2000metal,gibson1982mechanics,koudelka2012effective}, it can be conjectured that the adopted \emph{planar} beam representation has a limited capability in predicting the complex behavior of \Alporas{} foam. Possibly, it lacks the stiffness contribution from the out-of-plane beams and membranes as well as the cell walls parallel to the investigated plane, though, the membrane contribution of cell walls was reported negligible in the case of high-porosity foams~\cite{koudelka2012effective}. 

The qualitative analysis of the impact of the geometry representation clearly shows that the Voronoi mesh adopted in~\cite{nemecek2013twoscale} leads to the overestimated value of bulk modulus. Assuming only the volumetric deformation, the axial stiffness of straight beams dominates the behavior of the Voronoi model resulting in nearly incompressible behavior, whilst in the case of real-geometry and tiling-based meshes the axial and bending stiffness contribute equally. This explains the results reported in~\cite{nemecek2013twoscale} where the authors considered only the volumetric excitation. On top of that, based on their experimental observations, they assumed negligible Poisson's effect and equated Young's modulus to the oedometric component of the homogenized stiffness matrix. However, if the homogenization procedure was performed carefully, they would arrive at similar results as in \Tref{tab:foamresults}. It is worthwhile to note that zero Poisson's ratio is in contradiction also with the characteristics listed in~\Tref{tab:alporasproperties}, see~\cite{ashby2000metal} for further details. 

Despite the incapability of the two-dimensional beam model to capture behavior of complex materials such as highly porous metallic foams, the concept of Wang tilings proved to be a valuable technique to generate large computational domains of desired geometrical characteristics. Moreover, it allows to circumvent meshing of every single microstructure realization if the dicretization mesh is constructed on each tile separately such that it satisfies the given compatibility constrains. There is a promising possibility to extend the presented modeling strategy to three dimensions via Wang cubes, a spatial variant of Wang tiles, along with the spatial Voronoi tessellation to compress the true geometry of \Alporas{} foam, and to perform a similar analysis using shell elements instead of prismatic beams.

\section*{Acknowledgments}

The authors acknowledge the Czech Science Foundation, Grant No. 13-24027S. Special thanks belongs to Jan Zeman of CTU in Prague for substantial help in all aspects of the exposed topics. In addition, we would like to thank Ji\v{r}\'{i} and Michal \v{S}ejnoha, both of CTU in Prague, for valuable comments on the manuscript.


\section*{Bibliography}


\appendix
\section{Minimal kinematic boundary conditions for beam model}
\label{s:minimalKBC}

In the two-dimensional wireframe setting, \Eref{eq:minimal_condition} defining Minimal Kinematic Boundary Conditions can be rewritten into the following form
\begin{equation}
	\sevek{0} = \sum_{i=1}^{\nboundarybeams} \int_{L^{(i)}} \begin{Bmatrix} \scal{n}_1^{(i)} \scal{u}(x\loc) \\ \scal{n}_2^{(i)} \scal{w}(x\loc) \\ \scal{n}_1^{(i)} \scal{w}(x\loc) + \scal{n}_2^{(i)} \scal{u}(x\loc) \end{Bmatrix} \de{x\loc} \,,
	\label{eq:discreteminKBC}
\end{equation}
where $\nboundarybeams$ stands for the number of the boundary beams, $\scal{L}^{(i)}$ is the length of the $i$-th boundary beam, $\scal{n}_1^{(i)}$ and $\scal{n}_2^{(i)}$ are the components of the outward unit normal vector $\sevek{n}^{(i)}$, and $x\loc$ is the local coordinate measured along the beam axis.
\Eref{eq:discreteminKBC} allows to compute the $i$-th boundary beam contribution to the constraint matrix $\semtrx{C}^{\Cminkbc}$, \Eref{eq:boundary_constrain} as
\begin{equation}
\arraycolsep=2.5pt
	\semtrx{C}^{\Cminkbc,(i)} = \begin{bmatrix} \scal{n}_1^{(i)} & 0 \\ 0 & \scal{n}_2^{(i)} \\ \scal{n}_2^{(i)} & \scal{n}_1^{(i)} \end{bmatrix} \semtrx{T}^{\ell\mhyphen\mathrm{g}} \,  \semtrx{M} \, \semtrx{T}^{\mathrm{g}\mhyphen\ell}\,,
\end{equation}
where the matrices $\semtrx{T}^{\ell\mhyphen\mathrm{g}}$ and $\semtrx{T}^{\mathrm{g}\mhyphen\ell}$ transform the corresponding quantities between the local and the global coordinate systems. Namely,
\begin{equation}
\arraycolsep=2.5pt
	\semtrx{T}^{\ell\mhyphen\mathrm{g}} = \begin{bmatrix} \cos \alpha^{(i)} & -\sin \alpha^{(i)} \\ \sin \alpha^{(i)} & \cos \alpha^{(i)} \end{bmatrix}
\end{equation}
and
\begin{equation}
\arraycolsep=2.5pt
	\semtrx{T}^{\mathrm{g}\mhyphen\ell}\!=\!\begin{bmatrix} 
		\cos \alpha^{(i)} & \sin \alpha^{(i)} & 0 & 0 & 0 & 0 \\
		-\sin \alpha^{(i)} & \cos \alpha^{(i)} & 0 & 0 & 0 & 0 \\
		0 & 0 & 1 & 0 & 0 & 0 \\
		0 & 0 & 0 & \cos \alpha^{(i)} & \sin \alpha^{(i)} & 0 \\
		0 & 0 & 0 & -\sin \alpha^{(i)} & \cos \alpha^{(i)} & 0 \\
		0 & 0 & 0 & 0 & 0 & 1
	\end{bmatrix}\!\!,
\end{equation}
where $\alpha^{(i)}$ is the directional angle of the $i$-th beam. The matrix $\semtrx{M}$
\begin{equation}
\arraycolsep=2.5pt
	\semtrx{M} = \begin{bmatrix} \frac{L^{(i)}}{2} & 0 & 0 & \frac{L^{(i)}}{2} & 0 & 0 \\ 0 & \frac{L^{(i)}}{2} & \frac{{L^{(i)}}^2}{12} & 0 & \frac{L^{(i)}}{2} &  \frac{-{L^{(i)}}^2}{12} \end{bmatrix}
\label{eq:}
\end{equation}
is obtained by plugging the the shape functions for $u$ and $w$, e.g.~\cite{bittnar_numerical_1996},
\begin{equation}
\begin{split}
	N_{u}^{u_1}\left(\xi\right) &=  1 - \xi \,,\\
	N_{u}^{u_2}\left(\xi\right) &=  \xi \,,\\
	N_{w}^{w_1}\left(\xi\right) &= \frac{1}{1+2\kappa} \left[ (1+2\kappa) - 2\kappa\xi - 3\xi^2 + 2\xi^3 \right]\!,\\
	N_{w}^{\phi_1}\left(\xi\right) &= \frac{L}{1+2\kappa} \left[  -\left( 1+\kappa \right) \xi + \left( 2+\kappa \right) \xi^2 - \xi^3 \right]\!,\\
	N_{w}^{w_2}\left(\xi\right) &= \frac{1}{1+2\kappa} \left[ (2\kappa\xi + 3\xi^2 - 2\xi^3 \right]\!,\\
	N_{w}^{\phi_2}\left(\xi\right) &= \frac{L}{1+2\kappa} \left[  \kappa \xi + \left( 1-\kappa \right) \xi^2 - \xi^3 \right]\!,
\end{split}
\end{equation}
where $\xi$ is the normalized coordinate $\xi = \frac{x\loc}{\scal{L}^{(i)}}$, $\kappa = \frac{EI}{kGA}$ and $k$ is the Timoshenko shear coefficient, in the approximation of the local beam displacements
\begin{equation}
\renewcommand\arraystretch{1.0}
\arraycolsep=2.0pt
	\begin{Bmatrix} u\loc \\ w\loc \end{Bmatrix} 
	= 
	\begin{bmatrix} N_{u}^{u_1} & 0 & 0 & N_{u}^{u_2} & 0 & 0 \\ 0 & N_{w}^{w_1} & N_{w}^{\phi_1} & 0 & N_{w}^{w_2} & N_{w}^{\phi_2} \end{bmatrix}
	\begin{Bmatrix} \scal{u}_1\loc \\ \scal{w}_1\loc \\ \scal{\phi}_1\loc \\ \scal{u}_2\loc \\ \scal{w}_2\loc \\ \scal{\phi}_2\loc \end{Bmatrix}
\end{equation}
and evaluating the integral in~\Eref{eq:discreteminKBC}.


\section{Least square method}
\label{s:leastsquare}

The eigenvalues of the isotropic plane-strain stiffness matrix written as a linear combination of parameters $K$ and $G$ read as
\begin{equation}
  \sevek{\lambda}^{\textrm{iso}} = 
  \semtrx{A} \sevek{d} =
  \begin{bmatrix}
    1 & 0 \\
    2 & 0 \\
    \frac{2}{3} & 2
  \end{bmatrix}
  \begin{Bmatrix}
    G \\ K
  \end{Bmatrix} \,.
  \label{eq:KGrelation}
\end{equation}
The least square method applied to \Eref{eq:KGrelation} results in minimizing the distance between the calculated eigenvalues $\sevek{\lambda}^\homg$ of the homogenized stiffness matrix and $\sevek{\lambda}^{\textrm{iso}}$ in the Euclidean norm
\begin{equation}
  \norm{\sevek{\lambda}^{\textrm{iso}}  - \sevek{\lambda}^{\homg} }^{2} = \left(  \sevek{\lambda}^{\textrm{iso}}  - \sevek{\lambda}^{\homg} \right)^\trn \left(  \sevek{\lambda}^{\textrm{iso}}  - \sevek{\lambda}^{\homg} \right) \,.
  \label{eq:LSQnorm}
\end{equation}
Plugging \Eref{eq:KGrelation} into \Eref{eq:LSQnorm} and minimizing it with respect to $\sevek{d}$ yields the parameters $K^\homg$ and $G^\homg$ explicitly approximated as a projection of $\sevek{\lambda}^\homg$ onto the range of $\semtrx{A}$
\begin{equation}
\arraycolsep=2.5pt
  \sevek{d}^{\homg}\!= \left( \semtrx{A}^\trn \semtrx{A}\right)^{-1}\!\semtrx{A}^\trn \sevek{\lambda}^\homg\!=
  \begin{bmatrix} \frac{1}{5} & \frac{2}{5} & 0 \\ -\frac{2}{30} & -\frac{4}{30} & \frac{1}{2} \end{bmatrix} \sevek{\lambda}^\homg,
\label{eq:leastsquareKGApp}
\end{equation}
where $\left( \semtrx{A}^\trn \semtrx{A}\right)^{-1}\!\semtrx{A}^\trn$ is the left pseudoinverse of $\semtrx{A}$.

Note that applying the above procedure in plane-stress conditions leads to non-linear relation between the eigenvalues and the elastic parameters
\begin{equation}
  \sevek{\lambda}^{\textrm{iso}} = 
  \begin{Bmatrix}
    G & 2 G & \frac{18 K G}{4G + 3K}
  \end{Bmatrix}^\trn \,.
\end{equation} 
As a result, no explicit formula can be derived and an iterative Least square method must be used instead.

\end{document}